\documentclass[fleqn,usenatbib]{mnras}

\usepackage{newtxtext,newtxmath}
\usepackage[T1]{fontenc}

\DeclareRobustCommand{\VAN}[3]{#2}
\let\VANthebibliography\thebibliography
\def\thebibliography{\DeclareRobustCommand{\VAN}[3]{##3}\VANthebibliography}

\usepackage{xspace}
\usepackage{xcolor}
\definecolor{darkblue}{rgb}{0.0, 0.0, 0.55}
\definecolor{darkgreen}{rgb}{0.0, 0.55, 0.2}
\definecolor{darkred}{rgb}{0.55, 0.0, 0}
\usepackage{multirow}
\usepackage{mathtools}
\usepackage{bm}
\usepackage{comment}
\usepackage{graphicx}	% Including figure files
\usepackage{amsmath}	% Advanced maths commands
\newcommand{\lcdm}{$\Lambda$CDM\xspace}

\newcommand{\secref}[1]{\hyperref[#1]{section~\ref*{#1}}}
\newcommand{\appref}[1]{\hyperref[#1]{appendix~\ref*{#1}}}

\title[Cosmological parameters estimation for the SKAO from cross-correlation]{21cm Intensity Mapping cross-correlation with galaxy surveys:
current and forecasted cosmological parameters estimation for the SKAO}

\author[M. Berti et al.]{
Maria Berti,$^{1,2,3}$\thanks{E-mail: mberti@sissa.it}
Marta Spinelli,$^{4,5,6}$
and Matteo Viel$^{1,2,3,5,7}$
\\
$^{1}$SISSA- International School for Advanced Studies, Via Bonomea 265, 34136 Trieste, Italy\\
$^{2}$INFN – National Institute for Nuclear Physics, Via Valerio 2, 34127 Trieste, Italy\\
$^{3}$IFPU, Institute for Fundamental Physics of the Universe, via Beirut 2, 34151 Trieste, Italy\\
$^{4}$Institute for Particle Physics and Astrophysics,
ETH Z{\"u}rich, Wolfgang Pauli Strasse 27, 8093 Z{\"u}rich, Switzerland\\
$^{5}$INAF, Osservatorio Astronomico di Trieste, Via G. B. Tiepolo 11, I-34131 Trieste, Italy\\
$^{6}$Department of Physics and Astronomy, University of the Western Cape, Robert Sobukhwe Road, Bellville, 7535, South Africa\\
$^{7}$Italian Research Center on High Performance Computing, Big Data and Quantum Computing\\
}

\date{Accepted XXX. Received YYY; in original form ZZZ}

\pubyear{2022}

\begin{document}
\label{firstpage}
\pagerange{\pageref{firstpage}--\pageref{lastpage}}
\maketitle

% Abstract of the paper MAX 250 parole
\begin{abstract}
We present a comprehensive set of forecasts for the cross-correlation signal between 21cm intensity mapping and galaxy redshift surveys. We focus on the data sets that will be provided by the SKAO for the 21cm signal, DESI and Euclid for galaxy clustering. We build a likelihood which takes into account the effect of the beam for the radio observations, the Alcock-Paczynski effect, a simple parameterization of astrophysical nuisances, and fully exploit the tomographic power of such observations in the range $z=0.7-1.8$ at linear and mildly non-linear scales ($k<0.25 h/$Mpc). The forecasted constraints, obtained with Monte Carlo Markov Chains techniques in a Bayesian framework, in terms of the six base parameters of the standard $\Lambda$CDM model, are promising. The predicted signal-to-noise ratio for the cross-correlation
can reach $\sim 50$ for $z\sim 1$ and $k\sim 0.1 h/$ Mpc.
When the cross-correlation signal is combined with current Cosmic Microwave Background (CMB) data from Planck, the error bar on $\Omega_{\rm c}\,h^2$ and $H_0$ is reduced by a factor 3 and 6,  respectively, compared to CMB only data, due to the measurement of matter clustering provided by the two observables. The cross-correlation signal has a constraining power that is comparable to the auto-correlation one and combining all the clustering measurements a sub-percent error bar of 0.33\% on $H_0$ can be achieved, which is about a factor 2 better than CMB only measurements.
Finally, as a proof-of-concept, we test the full pipeline on the real data measured by the MeerKat collaboration \citep{Cunnington:2022uzo} presenting some (weak) constraints on cosmological parameters.
\end{abstract}


%%%%%%%%%%%%%%%%%%%%%%%%%%%%%%%%%%%%%%%%%%%%%%%%%%

%%%%%%%%%%%%%%%%% BODY OF PAPER %%%%%%%%%%%%%%%%%%

\section{Introduction}
Efforts to seek new physics beyond the standard cosmological model, which is grounded in cold dark matter and the cosmological constant ($\Lambda$CDM), are currently centred on addressing the $H_0$ and $S_8$ tensions. These involve modifications of the theoretical framework and the exploration of new observables with the potential to be sensitive to various scales and redshifts while being subject to distinct systematics and statistical errors compared to well-established probes.

\medskip
Neutral hydrogen (HI) has recently emerged as a new quantitative tracer of the large-scale structure (LSS) of the Universe \citep[e.g.][]{review,Ansari_2012,Santos:2015} via the intensity mapping (IM) technique.
The 21cm IM line is produced by the hyperfine structure spin-flip transition of the electron in atomic hydrogen \citep{Furlanetto:2006} and, compared to other observables, has the great advantage of probing large volumes in an efficient way at the expense of a relatively poor angular resolution.

Several planned and ongoing experiments, like compact interferometers (e.g. CHIME~\citep{Bandura:2014gwa,CHIMEdetection}, CHORD or HIRAX~\citep{Newburgh:2016mwi}) or single-dish telescopes (such as GBT~\citep{Masui2013,Wolz2022} or FAST~\citep{Hu:2019okh}) aim at measuring the IM signal ~\citep{Bharadwaj:2000,Battye:2004,McQuinn:2005,Chang:2007,Seo:2009fq,Kovetz:2017,Villaescusa-Navarro:2018} and some of them have achieved the detection of the HI signal in cross-correlation with galaxy surveys \citep{Chang2010,Masui2013,Anderson2018,Wolz2022}, since this measurement is likely to be less prone to systematics like foregrounds.

\medskip
The SKA Observatory (SKAO)\footnote{\url{https://www.skao.int/}}, consists of SKA-Low and
SKA-Mid telescopes, which will be located in Australia and South Africa, respectively. Using the SKA-Mid telescope array as a collection of single-dishes~\citep{Battye:2013,Santos:2015} it will be possible to perform 21cm IM at cosmological scales up to redshift $z\sim 3$~\citep{Bacon:2018}.  
 The SKAO is currently under construction and MeerKAT, the SKA-Mid precursor, has been conducting IM survey for cosmology~\citep[MeerKLASS]{Santos:2017}. Preliminary data analyses have provided interesting results { on the potential of the MeerKAT telescope intensity mapping surveys}~\citep{Wang:2021,irfan2022} along with a first {MeerKLASS} detection of the HI signal in cross-correlation with WiggleZ galaxies~\citep{Cunnington:2022uzo}. More recently, another breakthrough has been reached with the detection at cosmological scales of HI with IM in the auto power-spectrum ~\citep{paul23}.
 However, mitigating the foregrounds and their impact on the extracted signal, both in cross and auto-correlation, is challenging and several foreground removal techniques have been proposed ~\citep{Alonso2015,Wolz2016,Carucci2020,Matshawule:2021,Irfan2021,Cunnington2021,Soares:2020,Soares2021GPR,Spinelli:2021emp}. 

 \medskip
 From the theoretical point of view, it is of great importance to refine the forecasts for the 21cm IM, both alone and in combination with other probes, to optimise the survey design in order to enhance the signal-to-noise ratio \citep{villa15,bull:2015,Jiang:2023zex}, to address
 the non-linear scales modelling in the context of the MeerKat detection \citep{Padmanabhan:2023hfr}, to investigate the cross-correlation in the connection with galaxy formation models \citep{Spinelli:2019}, and to estimate the impact of fundamental new physics on the observables, like non-gaussianities or dark energy \citep{Jolicoeur:2023tcu,Wu:2022jkf}.

\medskip
{ In~\cite{Berti:2022ilk} we built on the formalism of~\citet{Blake:2019,Cunnington:2020, Soares:2020, Cunnington:2022uzo} and study the redshift-space 21cm power spectrum monopole and quadrupole, forecasting the constraining power of SKAO observations within multiple redshift bins. In this work, we extend this analysis by including in our pipeline the modelling of the 21cm and galaxy clustering cross-correlation signal. As before, we focus on the parameters of the $\Lambda$CDM model and exploit the exquisite tomographic nature of the 21cm IM signal.}

For the 21cm IM, we mimic SKA-Mid observations following the SKAO Redbook prescriptions \citep{Bacon:2018}.
Regarding galaxies, we rely on the mocking of data sets for the galaxy clustering signal which could be provided soon by Dark Energy Spectroscopic Survey (DESI)~\citep{Vargas-Magana:2018rbb,DESI:2016igz}
and the Euclid mission \citep{Euclid:2019clj}. We will build a full likelihood integrated within the \texttt{CosmoMC} code~\citep{Lewis:2002,Lewis:2013} and compute constraints through Markov-Chain Monte-Carlo (MCMC) techniques. We assess the constraining power of our mock 21cm data set in cross-correlation with galaxy clustering alone and combined with CMB data. 
 
We note that forecasts for future IM observations based on the Fisher matrix formalism have also been presented in \citet{obuljen18,Karagiannis:2022ylq,Viljoen:2020efi}. {Our work expands on previous studies in the following aspects: $i)$ we constrain the complete set of cosmological parameters, $ii)$ we cross-correlate the signal with the most recent forecasts for state-of-the-art galaxy surveys, $iii)$ we combine the 21cm data within multiple redshift bins with the Planck latest available results, $iv)$ we conduct complete MCMC analyses, to estimate the full posterior distribution. }

\medskip
 The structure of the paper is as follows. The methodology{, discussed more extensively in \cite{Berti:2022ilk},} is briefly reviewed in \secref{sec:methods}. The building of the mock observations is detailed in \secref{sec:cross_data_sets}. Results are discussed in \secref{sec:results_cross}, where we present the forecasted constraints obtained from 21cm and galaxy clustering cross-correlation alone (\secref{sec:res_cross}) and in combination with the latest Planck 2018 CMB data (\secref{sec:res_cross_planck}). Results obtained with data measured in \cite{Cunnington:2022uzo} are shown in \autoref{sec:res_real_data_constraints}. A summary of the results and our conclusions is outlined in \secref{sec:conclusions_cross}. We also provide two useful appendixes that discuss the Alcock-Paczynski effect's impact and a further cross-check on the attained signal on present data sets.

\section{Modeling the cross-correlation signal}
\label{sec:methods}
The analysis presented in this work for the 21cm $\times$ galaxy clustering power spectrum is an extension of the one discussed in \cite{Berti:2022ilk}. Since we adopt analogous formalism and framework of the previous study, in the following, we review only the essential information.
Having defined the cosmological model we consider in \secref{subsec:fiducial}, we discuss the 21cm auto-power spectrum and the galaxy power spectra in \secref{sec:P21_model_cross} and \secref{sec:galaxy_model}, which enter the error estimation. The model for the cross-correlation power spectrum and the power spectrum multipole expansion are presented in \secref{sec:cross_model} and \secref{sec:multipole_exp}. 

\subsection{Fiducial cosmological model}
\label{subsec:fiducial}
{
We work within the standard cosmological model framework, i.e. the $\Lambda$CDM model. We perform our analysis using the following six parameters to define the fiducial cosmology: $\Omega_bh^2$ and $\Omega_ch^2$, which describes the density of the baryonic and cold dark matter, respectively, the scalar spectral index $n_s$, the normalization of the primordial power spectrum $A_s$, the Thomson scattering optical depth due to reionization $\tau$, and  $\theta_{MC}$, that is connected to the angular scale of the sound horizon at decoupling. Moreover, we will focus on the derived parameters $H_0$, i.e. the current expansion rate in km s$^{-1}$ Mpc$^{-1}$ and $\sigma_8$, the root mean square matter fluctuations today in linear theory. 

Through all this work we assume a universe described by a Planck 2018~\citep{planck:2018} fiducial cosmology, i.e. $\{\Omega_b h^2 = 0.022383,\, \Omega_c h^2 = 0.12011,\, n_s= 0.96605 ,\, \ln (10^{10} A_s) =3.0448 ,\, \tau= 0.0543,\, H_0 = 67.32 \text{ km s$^{-1}$ Mpc$^{-1}$},\, \Sigma m_\nu = 0.06 \mathrm{eV} \}$, where $\Sigma m_\nu$ is the sum of neutrino masses in eV.
}

\subsection{Model for the observed 21cm signal power spectrum}
\label{sec:P21_model_cross}

The 21cm non-linear power spectrum can be modeled as~\citep{kaiser1987,Villaescusa-Navarro:2018, Bacon:2018}
\begin{equation}\label{eq:P21_cross}
     P_{21}(z,\,k,\, \mu) = \Bar{T}_{\rm b}^2(z) \left[ \left( b_{\mathrm{HI}}(z) + f(z)\, \mu^2\right)^2 P_{\rm m}(z,k) + P_{\rm{SN}}(z)\right],
\end{equation}
where $\bar{T}_{\rm b}$ is the {HI} mean brightness temperature, $b_{\mathrm{HI}}$ is the {HI} bias, $f$ is the growth rate, $\mu= \hat{k} \cdot \hat{z}$ is the cosine of the angle between the wave number and the line-of-sight, $P_{\rm m}(z,k)$ is the non-linear matter power spectrum and $P_{\rm{SN}}$ is the shot noise term.

For the evolution in redshift of the brightness temperature, we use the parametrization defined in \cite{Battye:2013}. Given that we lack an analytical model, $b_{\mathrm{HI}} (z)$ and $P_{\rm{SN}}(z)$ at a given redshift are computed by interpolating numerical results from hydrodynamical simulations~\citep{villa15,Villaescusa-Navarro:2018}. The growth rate $f(z)$ and the non-linear matter power spectrum $P_{\rm m}(z,k)$ are, instead, computed numerically by means of the Einstein-Boltzmann solver {\texttt{CAMB}}\footnote{See \url{https://CAMB.info/}. Note that non-linear corrections to the matter power spectrum are computed with the {{HALOFIT}}~\citep{Smith:2002} version from~\cite{Mead:2016}.}~\citep{lewis:2000}. 

\medskip
To mimic a realistic observation, we introduce the effect of a Gaussian telescope beam, as a suppression of the power spectrum on scales smaller than the beam's full width at half maximum \citep{Battye:2013,Villaescusa-Navarro:2016,Soares:2020,Cunnington:2020,Cunnington:2022}. The corresponding damping factor $\tilde{B} (z,k,\mu)$ can be written in terms of the beam's physical dimension $R_{\rm{beam}}$, as 
\begin{equation}
    \tilde{B} (z,k,\mu) = \exp \left[ \frac{-k^2 R_{\rm beam}^2(z) (1-\mu^2)}{2}\right].
\end{equation} 

In a real-world scenario, one must consider the possibility of having chosen the wrong fiducial cosmology. This can be taken into account with the Alcock–Paczynski ({AP}) modifications~\citep{Alcock:1979mp}. Anisotropies along the radial and transverse direction can be modelled as\footnote{In the literature, several definitions of $\alpha_\perp$ and $\alpha_\parallel$ have been proposed, e.g. \cite{Gil-Marin:2016wya,Hand:2017ilm,DAmico:2019fhj}. We follow the one of e.g.~\cite{Euclid:2019clj,Soares:2020}.}
\begin{equation}\label{eq:alcock_alphas}
    \alpha_\perp(z) = \frac{D_A(z)}{D_A^{\rm{fid}}(z)} \qquad \text{and} \qquad \alpha_\parallel(z) = \frac{H^{\rm{fid}}(z)}{H(z)}.
\end{equation} 
Here, $D_A^{\rm{fid}}(z)$ and $H^{\rm{fid}}(z)$ are the values of the angular diameter distance and the Hubble parameter at redshift $z$ predicted by the fiducial cosmology. The {AP} parameters $\alpha_\perp$ and $\alpha_\parallel$ modify the overall amplitude of the power spectrum and the wave vectors. The wave vector components along and transverse to the line of sight are then distorted as 
\begin{equation}
    q = \frac{k}{\alpha_\perp} \sqrt{1 + \mu^2 \left(\frac{\alpha_\perp^2}{\alpha_\parallel^2}-1\right)} 
\end{equation}
and 
\begin{equation}
    \nu = \frac{\alpha_\perp \mu }{\alpha_\parallel \sqrt{1 + \mu^2 \left(\frac{\alpha_\perp^2}{\alpha_\parallel^2}-1\right)}},
\end{equation}
where $k$ and $\mu$ are the assumed fiducial values of the wave vectors.

\medskip
The observed 21cm power spectrum, marked with the symbol $\hat{}$, including the beam smoothing, the {AP} effects and the instrumental noise, is then
\begin{equation}\label{eq:full_power_alcock}
     \Hat{P}_{21}(z,\,k,\, \mu) = \frac{1}{\alpha_\perp^2\alpha_\parallel}\tilde{B}^2 (z,q,\nu) P_{21}(z,q,\nu) { + P_{\rm{N}}}(z),
\end{equation}
where $P_{21}(z,q,\nu)$ is defined in~\autoref{eq:P21_cross}, but computed on the new variables $q$ and $\nu$. { The SKAO-like instrumental noise $P_{\rm N}$ can be modelled using the instrument specifications as in Equation 9 of \cite{Berti:2022ilk}.
}

We note that, in this {work}, we expand the modelling of \cite{Berti:2022ilk}, where we neglected the {AP} contribution in the first approximation. A discussion on the effect of the inclusion of the {AP} distortions on the cosmological parameter constraints is presented in {appendix} \ref{sec:appendix_AP}. 

\subsection{Model for the galaxy power spectrum}
\label{sec:galaxy_model}

The simplest parametrization of the galaxy power spectrum at a given redshift can be written as
\begin{equation}
     P_{\rm g}(z,k,\mu) = \left( b_{\mathrm{g}}(z) + f(z)\, \mu^2\right)^2 P_{\rm m}(z,k) + \frac{1}{\bar{n}_{\rm g}(z)},
\end{equation}
where $b_{\mathrm{g}}$ is the galaxy bias and $\bar{n}_{\rm g}$ is the galaxy number density. The term $1/\bar{n}_{\rm g}$ is the shot noise term for the galaxy power spectrum. In this work, we use values of $b_{\mathrm{g}}$ and $\bar{n}_{\rm g}$ in agreement with the official expected values for the planned galaxy surveys, as discussed in \secref{sec:data_set}. 

Due to the different observing techniques, the galaxy power spectrum is not affected by the beam correction. The {AP} distortions, instead, are the ones described in the previous section for $P_{21}$. Therefore, the observed galaxy power spectrum we consider is
\begin{equation}
    \label{eq:P_galaxy}
     \Hat{P}_{\rm g}(z,\,k,\, \mu) = \frac{1}{\alpha_\perp^2\alpha_\parallel} P_{\rm g}(z,q,\nu).
\end{equation}

\subsection{The cross-correlation signal power spectrum}
\label{sec:cross_model}

To predict the cross-correlation power spectrum between the 21cm signal and galaxy clustering, we use the following model (see e.g.~\cite{Cunnington:2022uzo, Casas:2022vik})
\begin{equation}
    {P}_{\rm{21,g}}(z,k,\mu) = \bar{T}_b \left( b_{\rm{HI}} + f\mu^2 \right)\left( b_{\rm g} + f\mu^2 \right) P_{\rm m} (z,k,\mu),
\end{equation}
where all the quantities appearing here are defined in the previous sections. In the expression above we do not make explicit the redshift dependence of the brightness temperature, the bias, and the growth rate for the sake of notation easiness. Moreover, it can be shown that the shot noise contribution for the cross-correlation power spectrum is negligible~\citep{Castorina:2016bfm,Villaescusa-Navarro:2018}

Taking into account the intensity mapping beam effect and the {AP} distortions, the observed cross-correlation signal becomes
\begin{equation}\label{eq:cross_power}
    \Hat{P}_{\rm{21,g}}(z,k,\mu) = \frac{1}{\alpha_\perp^2\alpha_\parallel} r \tilde{B}_{\perp} (z, q,\nu) {P}_{\rm{21,g}}(z, q,\nu),
\end{equation}
with $r$ being a cross-correlation coefficient {that accounts for unknown effects that may modify the theoretical estimate of the correlation degree}.\footnote{The definition of $r$ is not unique (see e.g. the discussion in~\cite{Cunnington:2022uzo}). In this work, we consider it an overall constant for simplicity, given that $r$ is considered as a nuisance parameter (\secref{sec:nuisances}).}

\subsection{Multipole expansion}
\label{sec:multipole_exp}
The non-isotropic power spectrum can be decomposed using Legendre polynomials  $\mathcal{L}_\ell(\mu)$. The coefficients of the expansion, i.e. the multipoles of the power spectrum, are given by
\begin{equation}
\label{eq:ell_P_cross}
    \Hat{P}_{\rm X, \ell} (z,k) = \frac{(2\ell + 1)}{2} \int_{-1}^{1} {\rm d}\mu\, \mathcal{L}_\ell(\mu) \Hat{P}_{\rm X}(z,k,\mu),
\end{equation}
with $\rm X$ being either the 21cm intensity mapping ($\rm X= 21$), the galaxy clustering ($\rm X= \rm g$) or their cross-term ($\rm X= 21,\rm g$).
In this work, we use the auto-power spectrum and cross-correlation monopoles, for which $\ell=0$ and $\mathcal{L}_0(\mu)=1$. In particular, we focus on forecasting the cross-correlation power spectrum monopole $\Hat{P}_{21,\rm g, 0} (z,k)$. In the following, for clarity of notation, we drop the subscript 0 and simply refer to the monopoles as $\Hat{P}_{\rm{21,g}}(z,k)$, $\Hat{P}_{\rm{g}}(z,k)$ and $\Hat{P}_{\rm{21}}(z,k)$. 

\section{Constructing the mock cross-correlation data}
\label{sec:cross_data_sets}
In this section, we construct mock data sets of cross-correlation measurements from future surveys. The 21{cm} and galaxy surveys we take into account are presented in \secref{sec:surveys}. 
Details on the construction of the synthetic data set and the analysis framework are given in \secref{sec:data_set} and \secref{sec:numerical_analysis_cross}.
\subsection{Survey specifications}
\label{sec:surveys}

\subsubsection{21cm intensity mapping}
The main focus of our analysis is the 21cm intensity mapping signal that can be measured with the {SKAO} telescope. 
We consider, in particular, a cosmological survey with the SKA-Mid telescope in single-dish mode, following \citet{Bacon:2018}. We assume a Wide Band~1 survey that covers a sky area of 20 000 deg$^2$ in the frequency range $0.35-1.05$ GHz (i.e. the redshift range $0.35-3$). The used SKAO specifications are summarized in \autoref{tab:all_specifics}.
\begin{table}
	\centering
	\caption{Assumed specifications for SKA-Mid Wide band 1~\citep{Bacon:2018}, DESI ELG~\citep{DESI:2016igz,Casas:2022vik}, and Euclid-like surveys~\citep{Euclid:2019clj}. For simplicity, we refer to SKA-Mid as SKAO, to DESI ELG as DESI, and to Euclid-like as Euclid. We collect the used effective redshifts $z$ and bin widths $\Delta z$, the galaxy biases $b_{\rm g}$ and number densities $\bar{n}_{\rm g}$, that we express in units of [$10^{-4} h^3$ Mpc$^{-3}$], the 21cm intensity mapping bias $b_{\rm HI}$, and the 21cm power spectrum shot noise $P_{\rm SN}$, in units of [($h^{-1}$ Mpc)$^3$].}
	\label{tab:all_specifics}
	\begin{tabular*}{\columnwidth}{l@{\hspace*{80pt}}c@{\hspace*{10pt}}} 
 \hline
    \multicolumn{2}{c}{SKAO}\\
   \hline
   Band frequency range &  0.35 - 1.05 GHz \\
 Corresponding redshift range &  0.35 - 3 \\
  Dish diameter $D_{\rm dish}$ & 15 [m]\\
\hline
\end{tabular*}
\begin{tabular*}{\columnwidth}{l@{\hspace*{100pt}}c@{\hspace*{20pt}}}
\\
\hline
 \multicolumn{2}{c}{SKAO $\times$ Euclid}\\
   \hline
   Observed redshift range &  0.9 - 1.8 \\
   Overlapping survey area & 10000 [deg$^2$]\\
   Corresponding $\Omega_{\rm sur}$ & 3.0 [sr]\\
   \hline
   \end{tabular*}
\begin{tabular*}{\columnwidth}{l@{\hspace*{20pt}}c@{\hspace*{40pt}}c@{\hspace*{40pt}}c@{\hspace*{40pt}}c@{\hspace*{40pt}}l} 
    $z$ & 1. & 1.2 & 1.4 & 1.65\\
  $\Delta z$ & 0.2 & 0.2 & 0.2 & 0.3 \\
   $b_{\rm g}$ & 1.46&1.61&1.75&1.9 \\
 $\bar{n}_{\rm g}$ & 6.86& 5.58& 4.21& 2.61 \\
{$b_{\rm{HI}}$ }& {1.49   } &  { 1.60 }&  {1.71 } & { 1.84 }\\
{$P_{\rm{SN}}$ } & {124 }& { 114} & {101 }& {  85.0}  \\
   \hline
	\end{tabular*}
\begin{tabular*}{\columnwidth}{l@{\hspace*{95pt}}c@{\hspace*{30pt}}}
\\
\hline
 \multicolumn{2}{c}{SKAO $\times$ DESI}\\
 \hline
 Observed redshift range &  0.7 - 1.7  \\
Overlapping survey area & 5000 [deg$^2$]\\
Corresponding $\Omega_{\rm sur}$ & 1.5 [sr]\\
   \hline
	\end{tabular*}
 \begin{tabular*}{\columnwidth}{l@{\hspace*{7pt}}c@{\hspace*{7pt}}c@{\hspace*{7pt}}c@{\hspace*{7pt}}c@{\hspace*{7pt}}c@{\hspace*{7pt}}c@{\hspace*{7pt}}c@{\hspace*{7pt}}c@{\hspace*{7pt}}c@{\hspace*{7pt}}c@{\hspace*{7pt}}} 
    $z$ & 0.75 & 0.85 & 0.95 & 1.05 & 1.15 & 1.25 & 1.35 & 1.45 & 1.55 & 1.65 \\
  $\Delta z$ & 0.1 & 0.1 & 0.1 & 0.1 & 0.1 & 0.1 & 0.1 & 0.1 & 0.1 & 0.1 \\
   $b_{\rm g}$ & 1.05&1.08&1.11&1.14&1.18&1.21&1.25&1.28&1.32&1.36 \\
 $\bar{n}_{\rm g}$ & 11.2 &  8.32 & 8.16& 5.14&
       4.49& 4.19& 1.57& 1.35&
       0.921 & 0.344 \\
       $b_{\rm{HI}}$ & 1.35 & 1.40& 1.46& 1.52& 1.57&  1.63&
 1.68& 1.73&  1.78& 1.84 \\
 $P_{\rm{SN}}$  & 132& 130 & 126 & 122 & 116 &
 111 & 105 & 98 & 91 & 85\\
   \hline
	\end{tabular*}
\end{table}

\subsubsection{Galaxy surveys}
We assume a Euclid-like { and a DESI-like} spectroscopic galaxy survey. {For Euclid,} following what has been proposed in~\cite{Euclid:2019clj}, we consider observations within four different redshift bins in the range of 0.9 - 1.8. The assumed values of the galaxy bias and number density computed at each effective redshift are presented in \autoref{tab:all_specifics}. 

To obtain a cross-correlation signal, one must take into account observations of the same portion of the sky. In agreement with other studies in the literature, we assume a 10 000 deg$^2$ overlapping patch of the sky observed by the {SKAO} and a Euclid-like survey. 

Hereafter, we simply refer to the Euclid survey, where is understood that a Euclid-like survey as the one described above is intended. 

\medskip
To construct cross-correlation measurements between the {SKAO} and {DESI}, we follow \cite{Casas:2022vik}. We focus on the {DESI} Emission Line Galaxies ({ELG}), as they probe a redshift range similar to the one covered by Euclid, i.e. 0.7 - 1.7, making easier a direct comparison between the two experiments.  In \autoref{tab:all_specifics}, we report the assumed values of the galaxy bias and number density at each effective redshift and we consider an overlapping area between {DESI} {ELG} and {SKAO} of 5 000 deg$^2$. The smaller area overlap with respect to a Euclid-like survey is forced by the different hemisphere locations of the two telescopes.

\begin{figure}
	\includegraphics[width=\columnwidth]{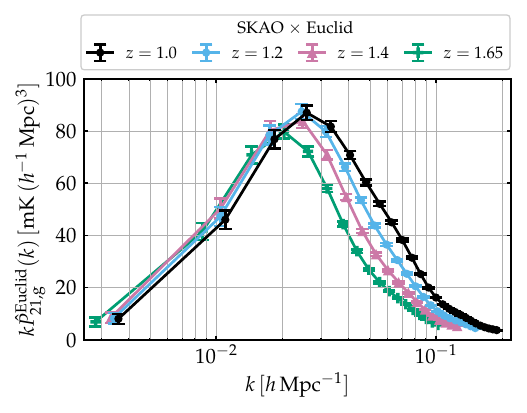}
	\includegraphics[width=\columnwidth]{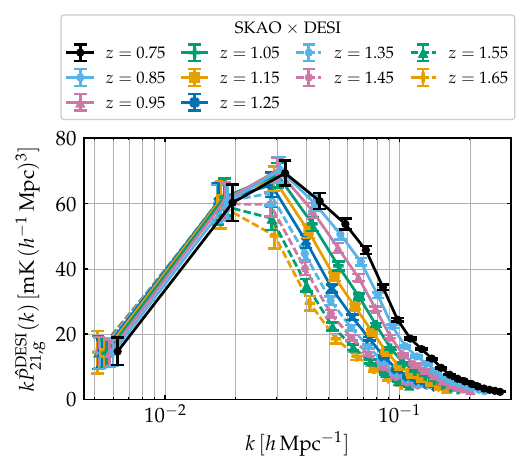}
    \caption{Mock data set for ${\text{SKAO}\times\text{Euclid}}$ (upper panel) and ${\text{SKAO}\times\text{DESI}}$ (lower panel) observations. The considered redshift bins are different for the two galaxy surveys. We refer to \secref{sec:cross_data_sets} for the extended discussion on how the signal and the errors are computed.}
    \label{fig:P_data_set_cross}
\end{figure}

\subsection{Mock data sets}
\label{sec:data_set}

We construct two different mock data sets for the 21cm and galaxy clustering cross-correlation power spectrum. One mimics an ${\text{SKAO}\times \text{Euclid}}$ analysis and the other a ${\text{SKAO}\times\text{DESI}}$ one, for the redshift bins and survey specifications described in \secref{sec:surveys} and \autoref{tab:all_specifics}. 

The scales that are accessible with the observations are fixed by the volume probed with the surveys in a given redshift bin. In Fourier space, the largest scale available at each effective redshift is $k_{\rm{min}}(z) = 2\pi / \sqrt[3]{V_{\rm{bin}}(z)}$, where $V_{\rm{bin}}(z)$ is the volume of each redshift bin, which we compute as 
\begin{align}\label{eq:volume_bin_cross}
\begin{split}
       V_{\rm{bin}}(z) &=  \Omega_{\rm{sur}} \int_{z - \Delta z /2}^{z + \Delta z /2} \mathrm{d}z' \, \frac{\mathrm{d} V}{\mathrm{d} z'\mathrm{d}\Omega} \\
       &=  \Omega_{\rm{sur}}  \int_{z - \Delta z /2}^{z + \Delta z /2} \mathrm{d}z' \, \frac{c r(z')^2}{H(z')}.
\end{split}
\end{align}
with $r(z)$ being the comoving distance and $\Omega_{\rm{sur}}$ the survey are in steradians (see \autoref{tab:all_specifics}). The smallest scale is, instead, imposed by the size of the {SKAO} telescope beam, due to the damping effect. It can be estimated as $k_{\rm{max}}(z) = 2\pi/R_{\rm{beam}}(z)$. At scales smaller than $k_{\rm{max}}$, the signal is dominated by the beam providing no relevant information on cosmology. Finally, we choose a fixed k-bin width as a function of redshift $\Delta k(z)$ in order to be large enough for modes to be independent, i.e. $\Delta k(z)\sim 2k_{\rm{min}}(z)$. 

\medskip
Assuming a set of $N$ measurements at redshift $z$ of the cross-correlation power spectrum $\Hat{P}_{\rm{21,g}}(k)$ at scales $\{k_1, \dots, k_N\}$, we estimate the error on at each point as (see e.g.~\citep{Smith09,Cunnington:2022uzo})
\begin{equation}\label{eq:errors}
    \hat{\sigma}_{\rm{21,g}}(k) = \frac{1}{\sqrt{2N_{\rm{modes}}(k)}} \sqrt{\hat{P}^2_{\mathrm{21,g}}(k) +  \hat{P}_{\mathrm{21}}(k) \hat{P}_{\rm g}(k) }, 
\end{equation}
where $\hat{P}_{\mathrm{21,g}}$ is the cross-correlation power spectrum defined in \autoref{eq:cross_power}, $\hat{P}_{\mathrm{21}}$ and $\hat{P}_{\rm g}$ are the 21cm and the galaxy power spectrum introduced in \autoref{eq:full_power_alcock} and \autoref{eq:P_galaxy} respectively.
Here, $N_{\rm{modes}}$ is the number of modes per $k$ and $\mu$ bin, computed as
\begin{equation}
     N_{\rm{modes}}(z,k) = \frac{k^2 \Delta k(z)}{4\pi^2} V_{\rm{bin}}(z).
\end{equation}
At each central redshift $z$ and data point $k$ we compute the cross-correlation power spectrum for ${\text{SKAO}\times \text{Euclid}}$ observations, labeled as $\hat{P}_{\mathrm{21,g}}^{\rm{Euclid}}(z,k)$, the one for ${\text{SKAO}\times \text{DESI}}$, $\hat{P}_{\mathrm{21,g}}^{\rm{DESI}}(z,k)$, and the corresponding errors, as discussed above. In table~\autoref{tab:all_specifics}, we gather some of the used redshift-dependent quantities. The resulting mock data sets are shown in \autoref{fig:P_data_set_cross}.

{ We highlight that in \autoref{eq:errors} the 21cm intensity mapping instrumental noise enters the estimate of the errors through $\hat{P}_{21}$. However, we find that the contribution of SKAO-like instrumental noise is minimal. This is not the case, instead, for a MeerKAT-like noise, which induces a non-negligible contribution to the error estimate, as in the analysis of \autoref{sec:res_real_data_constraints}. }

\subsection{Numerical analysis}
\label{sec:numerical_analysis_cross} 
In order to exploit the constraining power of the mock data set presented in \secref{sec:data_set}, we define a likelihood function and then set up the framework to constrain the cosmological parameters by adopting a Bayesian approach. Given a set of observations and a theory that depends on given parameters, the Bayes theorem links the posterior distribution to the likelihood function. The high-dimensional posterior can then be sampled using {MCMC} methods \citep[see e.g.]{gilks:1995}. {Following \cite{Berti:2022ilk}, we build a working pipeline to conduct full MCMC analyses on 21cm and cross-correlation observations. We test this pipeline by forecasting the constraining power of the datasets described above.}

\subsubsection{Likelihood function for the 21cm multipoles}
\label{sec:likelihood_cross}
Given a set of measurements at scales $\{k_1, \dots, k_N\}$ and redshift $z$, to compute the likelihood function we define the vector 
    ${\Theta}(z) = \Big( \Hat{P}_{\rm{21,g}}(z,k_1), \dots, \Hat{P}_{\rm{21,g}}(z,k_N) \Big)$.
The logarithmic likelihood is computed as
\begin{equation}
    \label{eq:likelihood_cross}
    -\ln\big[ \mathcal{L}\big] = \sum_{z} \,\frac{1}{2} \,\Delta {\Theta}(z)^{\rm T}\, {\mathsf{C}}^{-1}(z)\,  \Delta {\Theta}(z),
\end{equation}
where $\Delta {\Theta}(z) = {\Theta}^{\rm{th}}(z) - {\Theta}^{\rm{obs}}(z)$, the difference between the values of ${\Theta}(z)$ predicted from theory and observed. Here, ${\mathsf{C}}(z)$ is the covariance matrix computed as ${\mathsf{C}}(z) = \rm{diag}(\hat{\sigma}_{21, \rm g}^2(z,k_1),  \dots,\hat{\sigma}_{21, \rm g}^2(z,k_N) )$. We consider independent redshift bins, i.e. we simply sum over the contribution from each central redshift. 

\begin{figure}
	\includegraphics[width=\columnwidth]{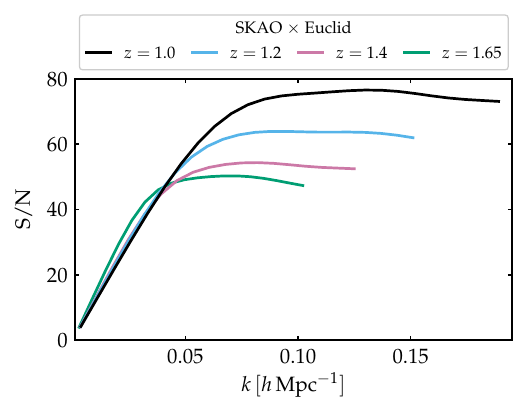}
	\includegraphics[width=\columnwidth]{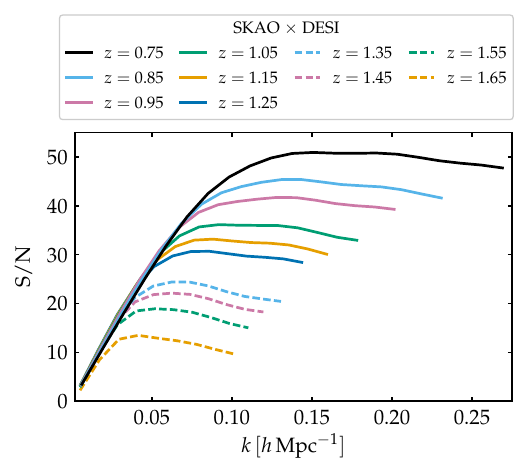}
    \caption{Predicted signal-to-noise ratio as a function of $k$ for ${\text{SKAO}\times\text{Euclid}}$ (upper panel) and ${\text{SKAO}\times\text{DESI}}$ (lower panel) mock observations. }
    \label{fig:signal_to_noise_cross}
\end{figure}

\medskip
\autoref{fig:signal_to_noise_cross} shows the signal-to-noise ratios as a function of $k$ in each redshift bin for both the constructed mock data sets. We observe that the signal-to-noise decreases at higher redshifts. The behaviour and orders of magnitude found here are compatible with the results for the 21{cm} power spectrum multipoles in~\cite{Soares:2020,Berti:2022ilk}. 

We conduct an {MCMC} analysis varying the six parameters describing the \lcdm model, i.e. we vary $\{\Omega_b h^2,\, \Omega_c h^2, n_s,\, \ln (10^{10} A_s),\, \tau,\, 100\theta_{\rm{MC}},\, \Sigma m_\nu, \, P_{{\rm{SN}},i} \}$ assuming wide flat priors on each of the parameters. Results on other parameters, such as $H_0$ and $\sigma_8$, are derived from results on this set. 
To perform the study, we develop a likelihood code integrated with the {MCMC} sampler \texttt{CosmoMC}\footnote{See \url{https://cosmologist.info/cosmomc}.}~\citep{Lewis:2002,Lewis:2013}. We further expand on the code we implemented and used in \cite{Berti:2021,Berti:2022ilk} including the computation of the theoretical expectations for the 21cm and galaxy clustering cross-correlation power spectrum and the relative likelihood function at different redshift. We recall that each redshift bin is considered independent, thus we consider a diagonal covariance matrix constructed with the forecasted errors. 

\subsubsection{Nuisance parameters}
\label{sec:nuisances}
As in \cite{Berti:2021,Berti:2022ilk}, along with the cosmological parameters we implement different nuisances. 
Indeed, the access to the matter clustering is not direct as it appears in \autoref{eq:cross_power} in combination with the brightness temperature and the {HI} bias and the galaxy bias. These quantities, although the scientific community hopes to obtain external measurements (e.g. the total neutral hydrogen density as a function of redshift, key unknown for the brightness temperature, is one of the scientific goals of the {MeerKAT} survey {Laduma}), may need to be treated as unconstrained quantities in a pessimistic scenario. 
To take into account this lack of knowledge, we allow for combinations of these parameters to vary in the {MCMC} run, thus leaving free the overall amplitude of the power spectrum. The contribution from the nuisances is then marginalised out in the final analysis.

To be completely agnostic, for the cross-correlation power spectrum we include in the nuisances also the correlation coefficient $r$ and the galaxy bias. Thus, we consider the following three combinations of parameters $\sqrt{r \bar{T}_b} b_{\rm{HI}} \sigma_8$, $\sqrt{r \bar{T}_b} b_{\rm{g}} \sigma_8$, and $\sqrt{r \bar{T}_b} f \sigma_8$, where we renormalized the matter power spectrum as $P_{\rm m}/\sigma_8^2$. 

\medskip
Given that all the parameters are redshift-dependent quantities, the actual number of nuisances is three times the number of redshift bins. This translates into 4 $\times$ 3 nuisance parameters for ${\text{SKAO}\times\text{Euclid}}$ and 10 $\times$ 3 for ${\text{SKAO}\times\text{DESI}}$. Especially in the latter case, the high number of parameters to vary can impact the numerical efficiency of the {MCMC} computations. Following what is already done in \cite{Berti:2022ilk}, for ${\text{SKAO}\times\text{DESI}}$ only we reduce the number of nuisances by constraining their redshift evolution through a polynomial parametrization. Rewriting $N(z) = az^3 + bz^2 + cz + d$ for $N(z) = \sqrt{r \bar{T}_b} b_{\rm{HI}} \sigma_8,\, \sqrt{r \bar{T}_b} b_{\rm{g}} \sigma_8,\, \sqrt{r \bar{T}_b} f \sigma_8$, we implement as nuisances the coefficient of the polynomial $a$, $b$, $c$, and $d$, reducing the number of nuisance parameters from 30 to 12. 

In the following, with the label "nuisances" or "nuis." we refer to the parameters described above. For each nuisance, we assume a wide flat prior { in the range $[-1,1]$}. 

\subsubsection{CMB likelihoods and data sets}
\label{sec:Planck_data_set}

In this study, we combine our mock 21cm and galaxy clustering cross-correlation data sets with Planck 2018~\citep{planck:2018}. The CMB likelihood includes the high-$\ell$ TT, TE, EE lite likelihood in the interval of multipoles $30\leq\ell\leq 2508$ for TT and $30\leq\ell\leq 19696$ for TE, EE. Lite likelihoods are calculated with the \texttt{Plik lite} likelihood~\citep{planck:2018like}. Instead for the low-$\ell$ TT power spectrum, we use data from the \texttt{Commander} component-separation algorithm in the range $2 \leq \ell \leq 29$. We also adopt the Planck CMB lensing likelihood and the low EE polarization power spectrum, referred to as lowE, in the range $2 \leq \ell\leq 29$, calculated from the likelihood code \texttt{SimAll}~\citep{planck:2018_maps}. In the rest of the paper with the label "Planck 2018" we refer to the combination TT, TE, EE + low-$\ell$ + lowE + lensing.

\section{Results}
\label{sec:results_cross}
We present in this section the results of our analysis. We first explore the constraining power of the mock cross-correlation data, with and without nuisances (\secref{sec:res_cross}). We then combine the mock data sets with Planck {CMB} data (\secref{sec:res_cross_planck}). Finally, in \secref{sec:res_real_data_constraints} we present the constraints we obtain on the cosmological parameters for the published measurement of the ${\text{MeerKAT}\times\text{WiggleZ}}$ cross-correlation power spectrum presented in~\cite{Cunnington:2022uzo}. 

\medskip
{As a reference}, throughout this analysis, we compare results from the cross-correlation forecast with the best results obtained with the 21cm multiples in {\cite{Berti:2022ilk}}, i.e. the fully non-linear monopole and quadrupole data set that we label as "$\Hat{P}_0 + \Hat{P}_2$". Note that we expand on the results of \cite{Berti:2022ilk} by introducing the {AP}, in order to be consistent with the modelling of the cross-correlation used in this {work}. Thus, "$\Hat{P}_0 + \Hat{P}_2$" here include {AP} effects. For a discussion on the impact of {AP} distortions, we refer to {appendix} \ref{sec:appendix_AP}. 

{We stress that when combining auto-power spectrum and cross-correlation forecasts, we neglect the covariance between the two data sets in the first approximation.}

We show the marginalised 1D and 2D posteriors for the studied set of parameters. Note that 68\% confidence level constraints are presented as percentages with respect to the marginalised mean value. 
\begin{figure}
	\includegraphics[width=\columnwidth]{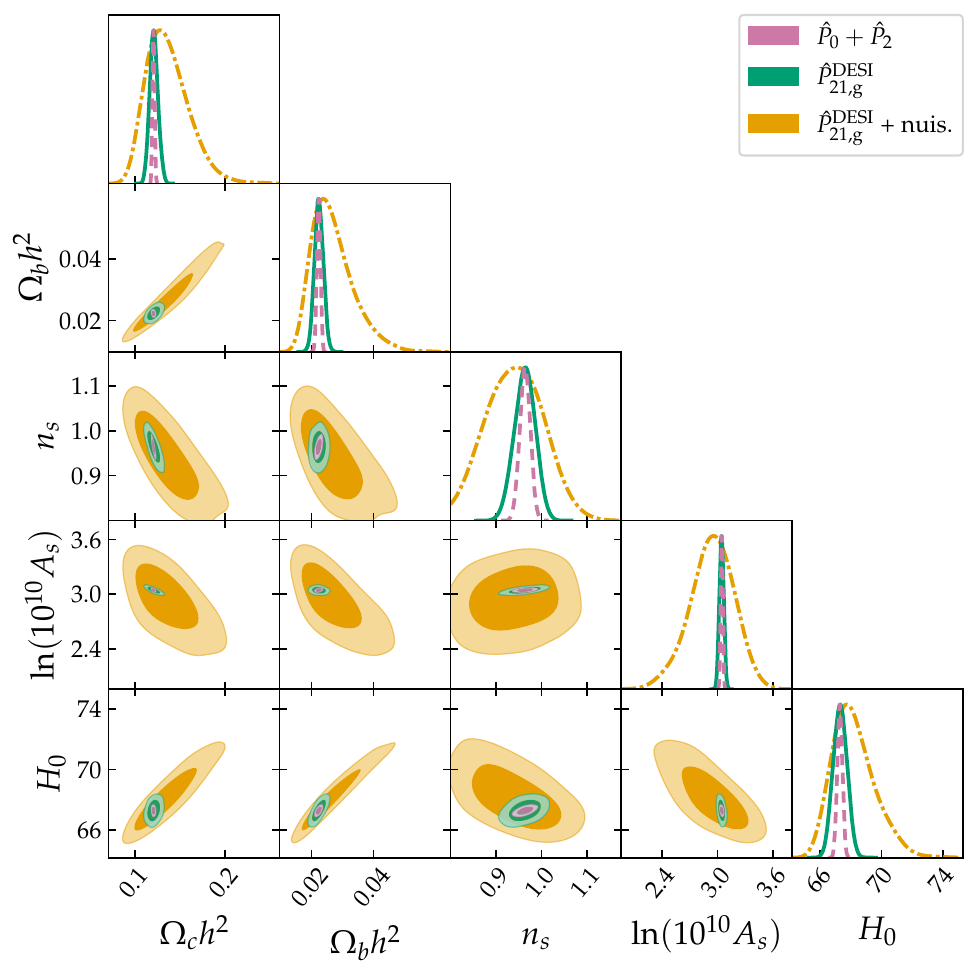}
    \caption{Joint constraints (68\% and 95\% confidence regions) and marginalised posterior distributions on a subset of the cosmological parameters. The label "$\Hat{P}_0 + \Hat{P}_2$" (dashed lines) stands for the forecasted 21cm power spectrum monopole and quadrupole observations (see~{appendix} \ref{sec:appendix_AP}). "$\hat{P}_{\mathrm{21,g}}^{\mathrm{DESI}}$" refers to the mock cross-correlation power spectrum data set constructed above. The label "nuis." (dashed-dotted lines) indicates that we vary the nuisance parameters along with the cosmological ones. The relative constraints are listed in \autoref{tab:constraints_alone}.}
    \label{fig:DESI_cross}
\end{figure}
\begin{figure}
	\includegraphics[width=\columnwidth]{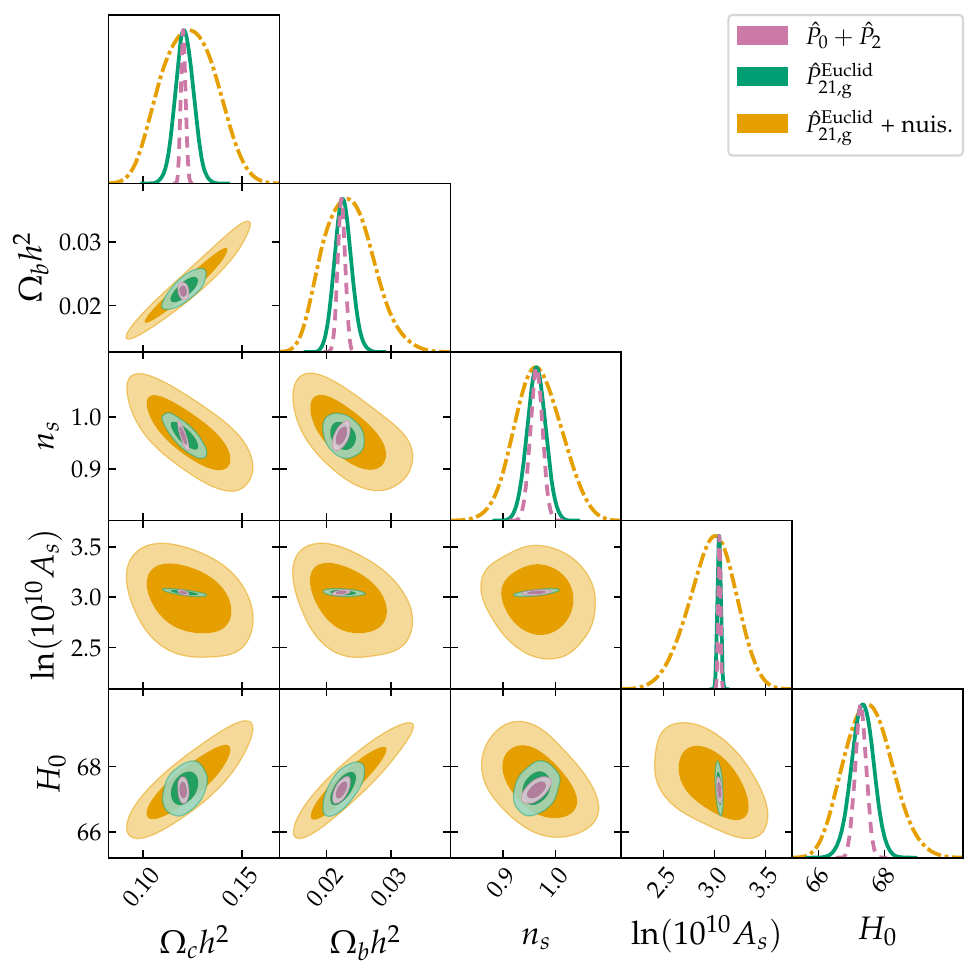}
    \caption{Joint constraints (68\% and 95\% confidence regions) and marginalised posterior distributions on a subset of the cosmological parameters. The label "$\Hat{P}_0 + \Hat{P}_2$" (dashed lines) stands for the forecasted 21cm power spectrum monopole and quadrupole observations (see~{appendix} \ref{sec:appendix_AP}). "$\hat{P}_{\mathrm{21,g}}^{\mathrm{Euclid}}$" refers to the mock cross-correlation power spectrum data set constructed above. The label "nuis." (dashed-dotted lines) indicates that we vary the nuisance parameters along with the cosmological ones. The relative constraints are listed in \autoref{tab:constraints_alone}.}
    \label{fig:Euclid_cross}
\end{figure}

\subsection{Probing the constraining power of future 21cm $\times$ galaxy clustering data}
\label{sec:res_cross}
\begin{table*}
	\centering
	\caption{Marginalised percentage constraints on cosmological parameters at the 68\% confidence level. We show the results obtained using different combinations of forecasted data sets. The label "$\Hat{P}_0 + \Hat{P}_2$" stands for the forecasted 21cm power spectrum monopole and quadrupole observations (see~{appendix} \ref{sec:appendix_AP}). "$\hat{P}_{\mathrm{21,g}}^{\mathrm{Euclid}}$" and "$\hat{P}_{\mathrm{21,g}}^{\mathrm{DESI}}$" refer to the mock cross-correlation power spectrum data sets constructed above. The label "nuis." indicates that we vary the nuisance parameters along with the cosmological ones. }
	\label{tab:constraints_alone}
	\begin{tabular}{lccccccc}
	\hline 
	\multirow{2}{*}{Parameter} & \multirow{2}{*}{$\hat{P}_0 + \hat{P}_2$} & \multirow{2}{*}{$\hat{P}_{\mathrm{21,g}}^{\mathrm{DESI}}$}& $\hat{P}_{\mathrm{21,g}}^{\mathrm{DESI}}$  &\multirow{2}{*}{$\hat{P}_{\mathrm{21,g}}^{\mathrm{Euclid}}$}& $\hat{P}_{\mathrm{21,g}}^{\mathrm{Euclid}}$  
 & $\hat{P}_{\mathrm{21,g}}^{\mathrm{DESI}}$ + $\hat{P}_{\mathrm{21,g}}^{\mathrm{Euclid}}$ &   $\hat{P}_0 + \hat{P}_2$  \\
 & & & + nuis. &  & + nuis.  & + nuis. & + $\hat{P}_{\mathrm{21,g}}^{\mathrm{DESI}}$ +$\hat{P}_{\mathrm{21,g}}^{\mathrm{Euclid}}$ + nuis. \\
		\hline
$\Omega_b h^2$ & 2.59\%& 6.43\%& 23.11\% & 5.78\%& 16.99\% & 12.52\%& 3.89\%\\
$\Omega_c h^2$ & 0.99\%& 3.81\%& 16.63\% & 3.75\%& 11.87\% &  8.59\%& 2.67\% \\
$n_s$ & 1.19\%& 2.43\%& 6.79\% & 1.82\%& 4.59\% & 3.56\%& 1.08\%  \\
${\rm{ln}}(10^{10} A_s)$ & 0.37\%& 0.78\%& 8.08\% & 0.54\%& 7.62\% &  4.73\%& 0.81\% \\
$100\theta_{MC}$ & 0.17\%& 0.39\%& 0.75\% & 0.30\%& 0.62\% &  0.54\%& 0.21\% \\\hline
$H_0$ & 0.25\%& 0.69\%& 1.96\% & 0.49\%& 1.07\% &  0.87\%& 0.33\% \\
$\sigma_8$ & 0.29\%& 0.40\%& 9.41\% & 0.58\%& 10.03\% & 6.37\%& 1.11\% \\
		\hline
	\end{tabular}
\end{table*}
\begin{figure*}
	\includegraphics[width=1.7\columnwidth]{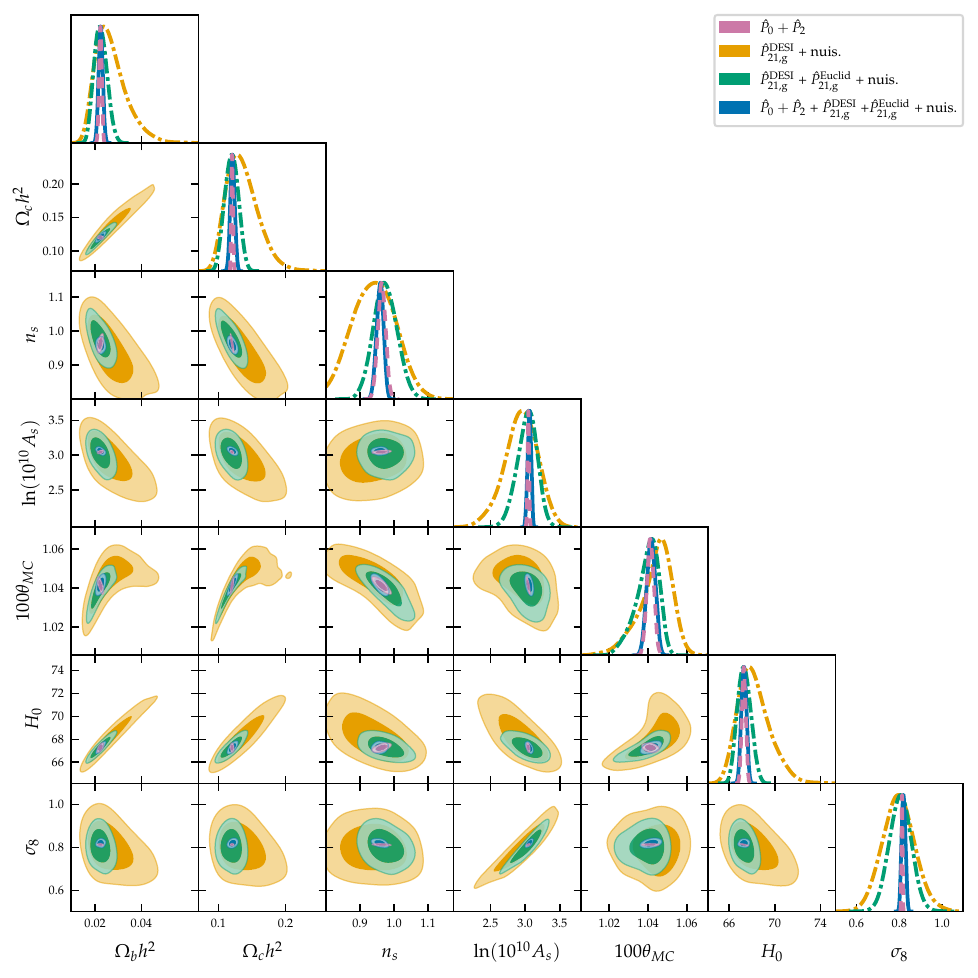}
    \caption{Joint constraints (68\% and 95\% confidence regions) and marginalised posterior distributions on a subset of the cosmological parameters. The label "$\Hat{P}_0 + \Hat{P}_2$" (dashed lines) stands for the forecasted 21cm power spectrum monopole and quadrupole observations (see~{appendix} \ref{sec:appendix_AP}). "$\hat{P}_{\mathrm{21,g}}^{\mathrm{Euclid}}$" and "$\hat{P}_{\mathrm{21,g}}^{\mathrm{DESI}}$" refer to the mock cross-correlation power spectrum data sets constructed above. The label "nuis." (dashed-dotted lines) indicates that we vary the nuisance parameters along with the cosmological ones. The relative constraints are listed in \autoref{tab:constraints_alone}.}
    \label{fig:full_P21_cross}
\end{figure*}

In \autoref{fig:DESI_cross} and \autoref{fig:Euclid_cross} we present the forecasted posterior distributions we obtain for the ${\text{SKAO}\times\text{DESI}}$ and ${\text{SKAO}\times\text{Euclid}}$ mock data sets we construct in this work. 
We show only some of the model parameters for brevity.

We obtain comparable results for both the ${\text{SKAO}\times\text{Euclid}}$ and the ${\text{SKAO}\times\text{DESI}}$ analysis. Looking at the 2D contours, we observe that the correlations between the cosmological parameters are similar and in line with the results obtained with the 21cm multipoles. The marked degeneracy in the $H_0 - \Omega_c h^2$ plane, found in previous works~\citep{Berti:2021, Berti:2022ilk}, is present also for the cross-correlation power spectrum case. As discussed in~\cite{Bardeen:1985tr}, measuring cosmological observables that strongly depend on the matter power spectrum, as $\hat{P}_{21,\rm g}$ does, fixes the shape of $P_{\rm m}$. This translates into fixing the quantity $\Omega_{\rm m} h$, which, in turn, induces the strong correlation $\Omega_c h^2 \propto H_0$. This feature is particularly relevant when combining 21cm observations with {CMB} data, as discussed in the next section.

As expected from the signal-to-noise estimates of \autoref{fig:signal_to_noise_cross}, better constraints are obtained for the ${\text{SKAO}\times\text{Euclid}}$ data set (\autoref{tab:constraints_alone}). Although {DESI} probes the same redshift range of Euclid and even with a higher number of redshift bins, Euclid will have a larger sky area of overlap with SKAO, suggesting that a larger sky coverage increases the constraining power more than the number of redshift bins. As expected, the best constraints are the ones obtained with the 21cm {multipoles}, in particular for $\Omega_c h^2$ and $H_0$. Indeed, the $\Hat{P}_0 + \Hat{P}_2$ is constructed to sample a wider redshift ($z=0 - 3$) and scales range (up to $k \sim 1\,h/$Mpc at low redshifts). It is interesting, however, to see that, despite these differences, cross-correlation results are still able to deliver competitive constraints. This makes a strong case for cross-correlation studies, especially in light of the reduced challenges in terms of residual systematics from the 21cm intensity mapping observations. 

\medskip
Adding the nuisance parameters, i.e. assuming no prior knowledge of the astrophysics at play, has the effect of varying the overall amplitude of the cross-correlation power spectrum. This translates into a broadening of the constraints, in particular on $A_s$. Moreover, the 2D contours are generally broader and show less clear correlations, except for the $H_0 - \Omega_c h^2$ and  $H_0 - \Omega_b h^2$ planes. Although the shape is stretched, the $H_0 - \Omega_c h^2$ degeneracy is still marked. 

\medskip
In \autoref{fig:full_P21_cross} we show the results on the full set of cosmological parameters for the combination of the two cross-correlation data sets and the 21cm multipoles one. Note that we do not report the constraints on $\tau$, due to the fact that the considered probes are not sensitive to this parameter. We compare the results with the ones for the 21cm multipoles and the $\hat{P}_{\mathrm{21,g}}^{\mathrm{DESI}}$ data set as a reference. In order to explore a more realistic scenario, we include the nuisance parameters. We observe that combining ${\text{SKAO}\times\text{DESI}}$ and ${\text{SKAO}\times\text{Euclid}}$ improves the constraints obtained with the two data sets separately. Including also the 21cm multipoles lead to the best result. With observations from 21cm probes only in the pessimistic case of including the nuisances, we are able to achieve constraints on the cosmological parameters comparable with Planck {CMB} observations. 

\medskip
We conclude that 21cm {IM} observations in cross-correlation with galaxy clustering seem to have a reduced constraining power with respect to 21cm auto-power spectrum measurements. However, when combined with the latter, they improve the constraints, showing that the cross-correlation signal carries complementary cosmological information. 

\subsection{Combining 21cm $\times$ galaxy clustering with {CMB} observations}
\label{sec:res_cross_planck}
Most recent forecast analyses find 21cm {IM} future observations to be a pivotal cosmological probe, highly complementary to {CMB} observations \citep{Bacon:2018}. Indeed, in \cite{Berti:2022ilk} we found that observations of the 21cm power spectrum multipoles contribute significantly to improving the constraints and reducing the degeneracies on the cosmological parameters. In this section, we investigate the effects of combining 21cm and galaxy clustering cross-correlations with {CMB} measurements.

For consistency, we first run the Planck likelihood in our framework and reproduce constraints in agreement with the Planck 2018 results. We then study the effect of adding the $\hat{P}_{\mathrm{21,g}}^{\mathrm{Euclid}}$ and $\hat{P}_{\mathrm{21,g}}^{\mathrm{DESI}}$ data sets and the two combined. As in \secref{sec:res_cross}, we compare the results we obtain with the constraints from the 21cm power spectrum {multipoles}. 
\begin{table*}
	\centering
	\caption{Marginalised percentage constraints on cosmological parameters at the 68\% confidence level. We show the results obtained using different combinations of forecasted data sets {in combination with Planck}. The label "Planck 2018" stands for TT, TE, EE + lowE + lensing, while the label "$\Hat{P}_0 + \Hat{P}_2$" refers to the forecasted 21cm power spectrum monopole and quadrupole observations (see~{appendix} \ref{sec:appendix_AP}). "$\hat{P}_{\mathrm{21,g}}^{\mathrm{Euclid}}$" and "$\hat{P}_{\mathrm{21,g}}^{\mathrm{DESI}}$" refer to the mock cross-correlation power spectrum data set constructed above. The label "nuis." indicates that we vary the nuisance parameters along with the cosmological ones.}
	\label{tab:constraints_planck}
	\begin{tabular}{lcccccccc} 
	\hline 
	\multirow{2}{*}{Parameter} & \multirow{2}{*}{Planck 2018} & 
 \multirow{2}{*}{+ $\hat{P}_0 + \hat{P}_2$ }& 
 \multirow{2}{*}{+ $\hat{P}_{\mathrm{21,g}}^{\mathrm{DESI}}$ }& 
 + $\hat{P}_{\mathrm{21,g}}^{\mathrm{DESI}}$  & 
 \multirow{2}{*}{+ $\hat{P}_{\mathrm{21,g}}^{\mathrm{Euclid}}$} & 
 + $\hat{P}_{\mathrm{21,g}}^{\mathrm{Euclid}}$  & 
  + $\hat{P}_{\mathrm{21,g}}^{\mathrm{DESI}}$ +$\hat{P}_{\mathrm{21,g}}^{\mathrm{Euclid}}$ & 
  +  $\hat{P}_0 + \hat{P}_2$ \\
 & & & & + nuis. & & + nuis.  & + nuis. &  + $\hat{P}_{\mathrm{21,g}}^{\mathrm{DESI}}$ +$\hat{P}_{\mathrm{21,g}}^{\mathrm{Euclid}}$ +nuis\\
		\hline
$\Omega_b h^2$ & 0.64\%& {0.47\% }& 0.48\%& 0.52\% & 0.48\%& 0.50\% &  0.49\%& 0.46\% \\
$\Omega_c h^2$ & 0.99\%& {0.26\% }& 0.39\%& 0.55\% & 0.36\%& 0.48\% &  0.42\%& 0.31\%  \\
$n_s$ & 0.42\%& {0.28\% }& 0.33\% & 0.34\% & {0.32\%} &0.32\% &0.32\%& 0.30\% \\
${\rm{ln}}(10^{10} A_s)$ & 0.46\%& {0.16\% }& 0.12\%& 0.44\% & 0.12\%& 0.45\% & 0.44\%& 0.43\% \\
$\tau$ & 13.44\%& {5.98\% }& 5.39\%& 12.09\% & 5.50\%& 11.82\% &  11.89\%& 11.38\%  \\
$100\theta_{\mathrm{MC}}$ & 0.03\% & {0.02\%}& 0.03\%& 0.03\% & 0.03\%& 0.02\% & 0.03\%& 0.03\%  \\\hline
$H_0$ & 0.79\%&{ 0.13\%}& 0.25\%& 0.41\% & 0.21\%& 0.33\% & 0.28\%& 0.14\% \\
$\sigma_8$ & 0.73\%&{ 0.24\% }& 0.24\%& 0.69\% & 0.24\%& 0.70\% & 0.69\%& 0.67\%\\
		\hline
	\end{tabular}
\end{table*}
\begin{figure*}
	\includegraphics[width=1.7\columnwidth]{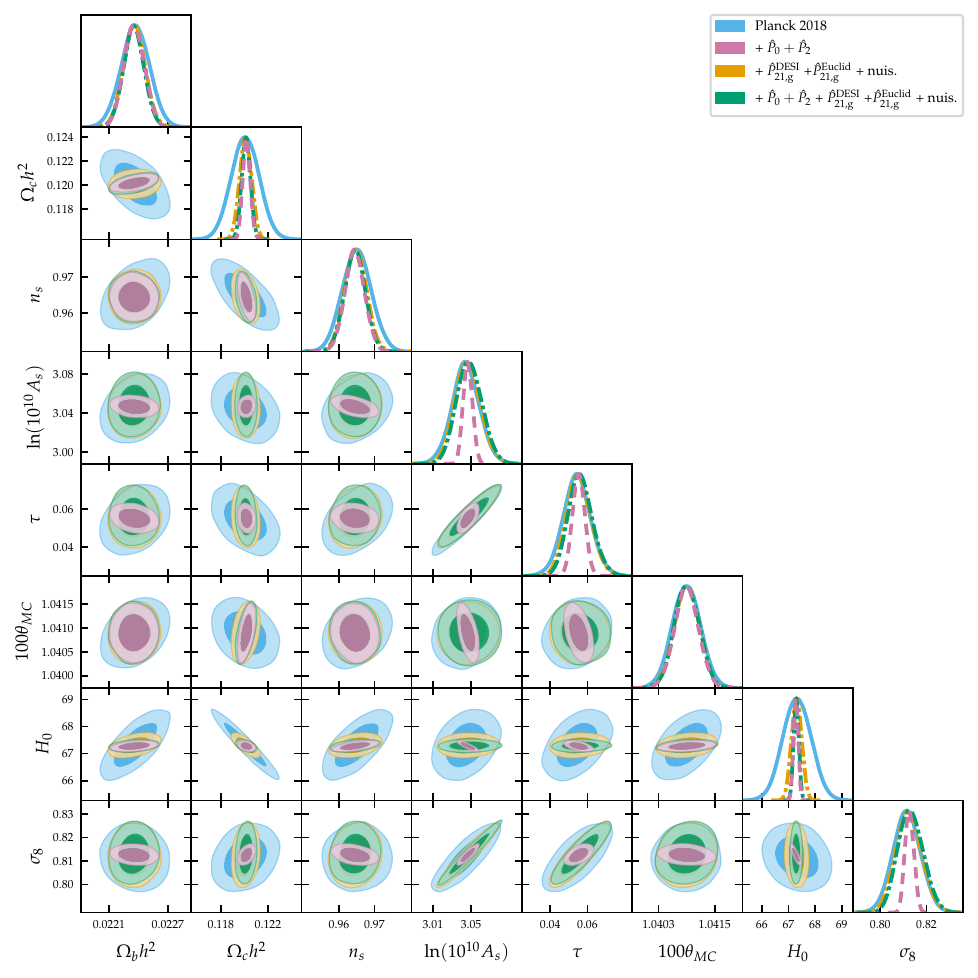}
    \caption{Joint constraints (68\% and 95\% confidence regions) and marginalised posterior distributions on a subset of the cosmological parameters. The label "Planck 2018" stands for TT, TE, EE + lowE + lensing, while the label "$\Hat{P}_0 + \Hat{P}_2$" (dashed lines) stands for the forecasted 21cm power spectrum monopole and quadrupole observations (see~{appendix} \ref{sec:appendix_AP}). "$\hat{P}_{\mathrm{21,g}}^{\mathrm{Euclid}}$" and "$\hat{P}_{\mathrm{21,g}}^{\mathrm{DESI}}$" refer to the mock cross-correlation power spectrum data sets constructed above. The label "nuis." (dashed-dotted lines) indicates that we vary the nuisance parameters along with the cosmological ones. The relative constraints are listed in \autoref{tab:constraints_planck}. }
    \label{fig:full_planck}
\end{figure*} 

\medskip
\autoref{tab:constraints_planck} shows the percentage constraints for this analysis. We observe that adding $\hat{P}_{\mathrm{21,g}}^{\mathrm{Euclid}}$ or $\hat{P}_{\mathrm{21,g}}^{\mathrm{DESI}}$ to Planck 2018 data reduces the estimated constraints with respect to the Planck alone results. The effect is prominent for $\Omega_ch^2 $ and $H_0$, for which the error is reduced by a factor of $\sim 3$, and $A_s$, with a factor of $\sim 4$ decrease. As one can see from \autoref{fig:full_planck}, in the $ H_0 - \Omega_ch^2$ plane the effect is ascribable to the correlation directions. Indeed, with Planck observations $ H_0$ and $\Omega_ch^2$ are anti-correlated, while they are positively correlated with $\hat{P}_{\mathrm{21,g}}$. Combining the two removes the degeneracy and reduces errors. The effect is also particularly evident for $A_{\rm s}$ since the {CMB} probes the quantity $A_{\rm s}\exp(-2\tau)$ and the matter power spectrum, which is constrained by 21cm data is sensitive to $S_8$, which is in turn degenerate with the optical depth to reionization as measured from the {CMB}. Therefore adding 21cm data effectively removes the degeneracies.

\medskip
When nuisance parameters are taken into account, as expected the improvement on the constraints is softened. In particular for $A_s$, and consequently $\sigma_8$, the constraining power is lost when the parameter space is open to the nuisances. Varying the nuisances corresponds to effectively changing the amplitude of the power spectrum and, thus, it results in worsened constraints on $A_s$. 

\medskip
The effects observed for the cross-correlation data sets combined with {CMB} are qualitatively comparable to the results obtained for the 21cm power spectrum multipoles. 
This confirms that when combining different kinds of 21cm observations with {CMB} data the improvement in the constraints is always driven mainly by the breaking of the degeneracy in the $\Omega_ch^2 - H_0$ plane. Indeed, our analysis reveals that even a less constraining measurement, such as the 21cm and galaxy cross-correlation, is effective in improving the errors on $\Omega_ch^2$ and $H_0$ if it presents a sufficiently marked correlation among these parameters.

To better prove this point, in \autoref{fig:full_planck} we compare the effect of combining Planck data with the 21cm multipoles, $\hat{P}_{\mathrm{21,g}}^{\mathrm{DESI}}$ and $\hat{P}_{\mathrm{21,g}}^{\mathrm{Euclid}}$, and the three combined. We find that even with the nuisances, results from $\hat{P}_{\mathrm{21,g}}^{\mathrm{DESI}}$ +$\hat{P}_{\mathrm{21,g}}^{\mathrm{Euclid}}$ (orange contours) are similar to the constraints from the {multipoles}, for which, instead, the nuisances are kept fixed as a best-result reference (pink contours). The main difference resides in the loss of constraining power on $A_s$ and the related parameters, which is however ascribable to the inclusion of the nuisances.
Further adding the 21cm {multipoles} to $\hat{P}_{\mathrm{21,g}}^{\mathrm{DESI}}$ +$\hat{P}_{\mathrm{21,g}}^{\mathrm{Euclid}}$ (green contours),\footnote{Note that in this case also the 21cm power spectrum nuisances are varied as in \cite{Berti:2022ilk}.} does not impact the constraining power or the shape of the correlations. This confirms that the 21cm probe is pivotal in breaking the {CMB} degeneracy in the $\Omega_ch^2 - H_0$ plane and the effect is relevant already at the level of cross-correlation or with {SKAO} precursor power spectrum measurements (see also \cite{Berti:2021}).

\medskip
We conclude that cross-correlations measurements of 21cm {IM} and galaxy clustering are a key cosmological probe complementary to {CMB} observations and, in combination with Planck, their forecasted constraining power is compatible with the one from 21cm power spectrum multipole measurements with the {SKAO}.

\subsection{State-of-the-art cosmological parameters constraints from the ${\text{MeerKAT}\times\text{WiggleZ}}$ detection}
\label{sec:res_real_data_constraints}
\begin{figure}
	\includegraphics[width=\columnwidth]{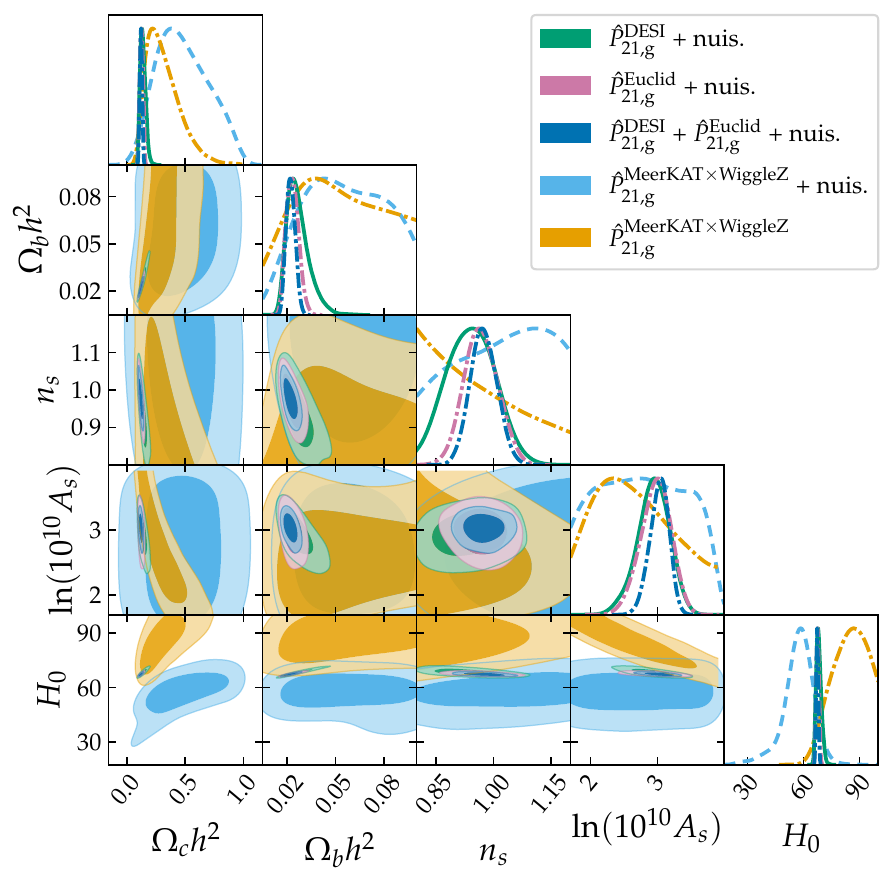}
    \caption{Joint constraints (68\% and 95\% confidence regions) and marginalised posterior distributions on a subset of the cosmological parameters. The labels "$\hat{P}_{\mathrm{21,g}}^{\mathrm{Euclid}}$" and "$\hat{P}_{\mathrm{21,g}}^{\mathrm{DESI}}$" refer to the mock cross-correlation power spectrum data sets constructed above. $\hat{P}_{\mathrm{21,g}}^{\mathrm{MeerKAT}\times\mathrm{WiggleZ}}$ refers to the cross-correlation power spectrum detection.
    The label "nuis." (dashed-dotted lines) indicates that we vary the nuisance parameters along with the cosmological ones. The relative constraints are listed in \autoref{tab:constraints_MeerKAT}.}
    \label{fig:real_data}
\end{figure}
\begin{table}
	\centering
	\caption{Marginalised percentage constraints on cosmological parameters at the 68\% confidence level. We show the constraints on the cosmological parameters obtained with the ${\text{MeerKAT}\times\text{WiggleZ}}$ cross-correlation power spectrum detection. The label "nuis." indicates that we vary the nuisance parameters along with the cosmological ones. The symbol "---" stands for unconstrained.}
	\label{tab:constraints_MeerKAT}
	\begin{tabular}{lcc} 
	\hline 
	{Parameter}  & $\hat{P}_{\mathrm{21,g}}^{\mathrm{MeerKAT}\times\mathrm{WiggleZ}}$& $\hat{P}_{\mathrm{21,g}}^{\mathrm{MeerKAT}\times\mathrm{WiggleZ}}$ + nuis. \\
		\hline
$ \Omega_b h^2 $  & --- & ---   \\
$ \Omega_c h^2 $  & $0.314^{+0.079}_{-0.18}$ & $0.48^{+0.19}_{-0.27}$   \\
$ 100\theta_{MC} $  & $1.090^{+0.061}_{-0.083}$ & $1.051^{+0.085}_{-0.070}$   \\
$ n_s $  & --- & ---  \\
$ {\rm{ln}}(10^{10} A_s) $  & --- & ---   \\
$\tau$ & --- & ---   \\\hline
$ H_0 $  & $84^{+10}_{-7}$ & $57^{+8}_{-6}$   \\
$ \sigma_8 $  & $0.974^{+0.068}_{-0.092}$ & $1.04^{+0.30}_{-0.49}$   \\\hline
	\end{tabular}
\end{table}
Cosmological 21cm observations with the {SKAO} will be possible in the upcoming years when the Observatory will be fully operational. However, the SKA-Mid pathfinder, MeerKAT is already taking data and its first cosmological surveys are promising. Recently, a power spectrum detection with the MeerKLSS survey, the intensity mapping survey with MeerKAT, in cross-correlation with galaxy clustering data has been made at the $7.7\sigma$ level~\citep{Cunnington:2022uzo}. The analysis pipeline we develop in this work is constructed to be ready to use with real cross-correlation power spectrum measurements. Therefore, we decide to test our methodology on the published results available for MeerKLASS. In the following, we present the result we obtain on the cosmological parameters constraints. We refer the interested reader to {appendix} \ref{app:test} for the technical consistency checks we run on the adopted power spectrum model and the predicted signal-to-noise ratio. 

\medskip
We tune the parameters of the likelihood function to match the settings of the observed data. Instead of the {SKAO} specifications, we use the MeerKLASS survey parameters, i.e. we consider a 200 dg$^2$ survey area and dishes of a diameter of $D_{\rm{dish}} = 13.5$ m. The observed effective redshift is $z=0.43$ with a bin width of $\Delta z = 0.059$. The signal is observed in cross-correlation with the WiggleZ 11h galaxy survey~\citep{Drinkwater2010:wigglez,WiggleZ:2018def}. When we do not include the nuisance parameters, we use the measured galaxy bias value $b_{\rm g} = 0.911$~\citep{Blake:2011rj} and cross-correlation factor $r=0.9$~\citep{Khandai:2010hs}. Other parameters and theoretical predictions are left unchanged. 

\medskip
We present the cosmological parameters constraints resulting from our {MCMC} analysis in \autoref{fig:real_data} and \autoref{tab:constraints_MeerKAT}. We observe that the state-of-the-art constraining power is limited with respect to the results forecasted for ${\text{SKAO}\times\text{Euclid}}$ and ${\text{SKAO}\times\text{DESI}}$, as expected due to the wider redshift ranges, probed scales, and survey area. Single bin MeerKAT observations are not yet able to constrain the complete set of cosmological parameters. However, the degeneracies between the parameters match the ones expected from our forecasts. In particular, the $H_0 - \Omega_c h^2$ correlation is clearly visible, although much less prominent. From these real measurements, we can infer new information on the marginalised mean value of the cosmological parameters. We find that all the constraints are compatible with the Planck results. 

The most constrained parameters are $\Omega_ch^2$ and $H_0$, proving that 21cm observations will be most useful to constrain them and their derived parameters. When fixing the nuisances, we find a high central value for $H_0$, although with a large error. We believe that this is not a physical effect, but is rather coming from a mismatch between the assumed brightness temperature value in our analysis and the one that seems to describe the observed data (see {appendix} \ref{app:test} for a more in-depth discussion). The conservative result is then the one in which nuisances are taken into account. In this case, we find that ${\text{MeerKAT}\times\text{WiggleZ}}$ data prefer a lower value of $H_0$, although the significance is not high enough to draw firm conclusions. 

From the constraints on the nuisances, one could estimate the value of $\Omega_{\rm{HI}}$. With our analysis we find the constraints on the nuisance parameters from real data to be too wide to infer a meaningful result.

Lastly, although we do not show here the results, we test the effect of combining ${\text{MeerKAT}\times\text{WiggleZ}}$ data with Planck 2018 observations. We find that the measured cross-correlation power spectrum does not increase significantly the constraining power of {CMB} observations, leaving the constraints and the 2D contours mostly unchanged.

\medskip
Although the constraining power of real detection is not yet competitive with other probes, the quality of the current 21cm {IM} observations in cross-correlation with galaxy clustering will improve sharply in the upcoming years and will soon become a useful independent cosmological probe. Moreover, the forecasted results for future surveys are very promising.

\section{Conclusions}
\label{sec:conclusions_cross}
In this work, we forecast the constraints on the \lcdm cosmological parameters 
for power spectrum measurements of 21cm intensity mapping in cross-correlation with galaxy clustering. Modelling the cross-power spectrum as in \cite{Cunnington:2022uzo}, we forecast mock observations of the {SKAO} cross-correlated with {DESI} and Euclid-like surveys. We test the constraining power of such probes alone and combined with the latest Planck {CMB} observations. Note that our modelling does not include possible residual foreground and systematics contamination.

We follow the {SKAO} Red Book~\citep{Bacon:2018} proposal and simulate single-dish observations with the SKA-Mid telescope in Band 1 (frequency range $0.35-1.05$ GHz). We cross-correlate this signal with a Euclid-like spectroscopic survey~\citep{Euclid:2019clj} and the {DESI} Emission Line Galaxies one~\citep{desiemission,Casas:2022vik} in the redshift range 0.7 - 1.7. Assuming a Planck 2018 fiducial cosmology, we construct two data sets of observations within multipole redshift bins. To test the constraining power on the cosmological parameters of our mock observations, we implement a likelihood function for the cross-correlation power spectrum, fully integrated with the {MCMC} sampler \texttt{CosmoMC}. We include a discussion on the impact of our lack of knowledge on the baryonic physics involved in the computation of the 21cm power spectrum as nuisance parameters in the analysis.

\medskip
We first focus on assessing the constraining power of cross-correlation observations alone, compared to the results we obtain for the 21cm multipoles. We, then, combine the two to investigate if they carry complementary information. The results of our analysis can be summarized as follows. 

\medskip
We find that {SKAO} power spectrum measurements in cross-correlation with galaxy clustering have a constraining power comparable to the 21cm auto-power spectrum. The ${\text{SKAO}\times\text{DESI}}$ and ${\text{SKAO}\times\text{Euclid}}$ data sets we construct are able to constrain the cosmological parameters up to the sub-percent level. They seem to be particularly effective on $H_0$, on which we obtain constraints between the $0.49 \%$ and the $1.96\%$ from 21cm and galaxy clustering cross-correlation alone. The tightest constraints are achieved when we combine 21cm power spectrum multipoles with the cross-correlation mock observations, for which we obtain a $0.33\%$ constraint on $H_0$, a value that is competitive with Planck. 

\medskip
When combining the cross-correlation mock measurements with {CMB} data, we find that they are pivotal to reducing the errors on the cosmological parameters. The effect is particularly prominent for $\Omega_ch^2$ and $H_0$, for which the errors are reduced by a factor between 2.5 - 1.8 and 3.8 - 2 respectively. Again, the best result is obtained by combining all the 21cm probes together. In this case, the error with respect to Planck alone results is reduced by a factor of 3.2 for $\Omega_c h^2$ and 5.6 for $H_0$, with the nuisance parameters taken into account. 

\medskip
Lastly, we test our analysis pipeline on the recent data for the cross-power spectrum between MeerKAT, the SKA-Mid pathfinder, and WiggleZ galaxy clustering \cite{Cunnington:2022uzo}. We find that state-of-the-art observations have limited constraining power on the complete set of cosmological parameters. However, the main features of the marginalised constraints are compatible with the forecasted results of this work. 

\medskip
{To conclude, our analysis supports the case of 21cm and galaxy clustering cross-correlation measurements. In combination with CMB observations, cross-correlations will be able to provide competitive constraints on the cosmological parameters comparable to the ones obtained with the auto-power spectrum.} The working pipeline presented in this work is found to be compatible and easily employable with real observations. The analysis we carry out can be straightforwardly adapted to forecast constraints on the neutrino mass and beyond \lcdm models. These extensions are currently under study.

\section*{Acknowledgements}

The authors would like to thank Steven Cunnington, Stefano Camera, José Fonseca, Alkistis Pourtsidou, and Laura Wolz for useful discussion and feedback.
MB and MV are supported by the INFN INDARK PD51 grant. MV acknowledges contribution from the agreement ASI-INAF n.2017-14-H.0.
MS acknowledges support from the AstroSignals Synergia grant CRSII5\_193826 from the Swiss National Science Foundation.

%%%%%%%%%%%%%%%%%%%%%%%%%%%%%%%%%%%%%%%%%%%%%%%%%%
\section*{Data Availability}

Access to the original code is available upon reasonable request to the corresponding author.

%%%%%%%%%%%%%%%%%%%% REFERENCES %%%%%%%%%%%%%%%%%%

\bibliographystyle{mnras}
\bibliography{Bibliography} 

\begin{thebibliography}{}
\makeatletter
\relax
\def\mn@urlcharsother{\let\do\@makeother \do\$\do\&\do\#\do\^\do\_\do\%\do\~}
\def\mn@doi{\begingroup\mn@urlcharsother \@ifnextchar [ {\mn@doi@}
  {\mn@doi@[]}}
\def\mn@doi@[#1]#2{\def\@tempa{#1}\ifx\@tempa\@empty \href
  {http://dx.doi.org/#2} {doi:#2}\else \href {http://dx.doi.org/#2} {#1}\fi
  \endgroup}
\def\mn@eprint#1#2{\mn@eprint@#1:#2::\@nil}
\def\mn@eprint@arXiv#1{\href {http://arxiv.org/abs/#1} {{\tt arXiv:#1}}}
\def\mn@eprint@dblp#1{\href {http://dblp.uni-trier.de/rec/bibtex/#1.xml}
  {dblp:#1}}
\def\mn@eprint@#1:#2:#3:#4\@nil{\def\@tempa {#1}\def\@tempb {#2}\def\@tempc
  {#3}\ifx \@tempc \@empty \let \@tempc \@tempb \let \@tempb \@tempa \fi \ifx
  \@tempb \@empty \def\@tempb {arXiv}\fi \@ifundefined
  {mn@eprint@\@tempb}{\@tempb:\@tempc}{\expandafter \expandafter \csname
  mn@eprint@\@tempb\endcsname \expandafter{\@tempc}}}

\bibitem[\protect\citeauthoryear{Aghamousa et~al.}{Aghamousa
  et~al.}{2016}]{DESI:2016igz}
Aghamousa A.,  et~al., 2016, {The DESI Experiment Part II: Instrument Design}
  (\mn@eprint {arXiv} {1611.00037})

\bibitem[\protect\citeauthoryear{Alcock \& Paczynski}{Alcock \&
  Paczynski}{1979}]{Alcock:1979mp}
Alcock C.,  Paczynski B.,  1979, \mn@doi [Nature] {10.1038/281358a0}, 281, 358

\bibitem[\protect\citeauthoryear{Alonso, Bull, Ferreira, Maartens  \&
  Santos}{Alonso et~al.}{2015}]{Alonso2015}
Alonso D.,  Bull P.,  Ferreira P.~G.,  Maartens R.,   Santos M.,  2015, \mn@doi
  [Astrophys. J.] {10.1088/0004-637X/814/2/145}, 814, 145

\bibitem[\protect\citeauthoryear{Amiri et~al.}{Amiri
  et~al.}{2023}]{CHIMEdetection}
Amiri M.,  et~al., 2023, \mn@doi [Astrophys. J.] {10.3847/1538-4357/acb13f},
  947, 16

\bibitem[\protect\citeauthoryear{{Anderson} et~al.,}{{Anderson}
  et~al.}{2018}]{Anderson2018}
{Anderson} C.~J.,  et~al., 2018, \mn@doi [\mnras] {10.1093/mnras/sty346}, \href
  {https://ui.adsabs.harvard.edu/abs/2018MNRAS.476.3382A} {476, 3382}

\bibitem[\protect\citeauthoryear{Ansari et~al.,}{Ansari
  et~al.}{2012}]{Ansari_2012}
Ansari R.,  et~al., 2012, \mn@doi [Astronomy {\&}Astrophysics]
  {10.1051/0004-6361/201117837}, 540, A129

\bibitem[\protect\citeauthoryear{Bandura et~al.}{Bandura
  et~al.}{2014}]{Bandura:2014gwa}
Bandura K.,  et~al., 2014, \mn@doi [Proc. SPIE Int. Soc. Opt. Eng.]
  {10.1117/12.2054950}, 9145, 22

\bibitem[\protect\citeauthoryear{Bardeen, Bond, Kaiser  \& Szalay}{Bardeen
  et~al.}{1986}]{Bardeen:1985tr}
Bardeen J.~M.,  Bond J.~R.,  Kaiser N.,   Szalay A.~S.,  1986, \mn@doi
  [Astrophys. J.] {10.1086/164143}, 304, 15

\bibitem[\protect\citeauthoryear{Battye, Davies  \& Weller}{Battye
  et~al.}{2004}]{Battye:2004}
Battye R.~A.,  Davies R.~D.,   Weller J.,  2004, \mn@doi [Mon. Not. Roy.
  Astron. Soc.] {10.1111/j.1365-2966.2004.08416.x}, 355, 1339

\bibitem[\protect\citeauthoryear{Battye, Browne, Dickinson, Heron, Maffei  \&
  Pourtsidou}{Battye et~al.}{2013}]{Battye:2013}
Battye R.~A.,  Browne I. W.~A.,  Dickinson C.,  Heron G.,  Maffei B.,
  Pourtsidou A.,  2013, \mn@doi [Mon. Not. Roy. Astron. Soc.]
  {10.1093/mnras/stt1082}, 434, 1239

\bibitem[\protect\citeauthoryear{Bernal, Breysse, Gil-Mar\'\i{}n  \&
  Kovetz}{Bernal et~al.}{2019}]{Bernal:2019}
Bernal J.~L.,  Breysse P.~C.,  Gil-Mar\'\i{}n H.,   Kovetz E.~D.,  2019,
  \mn@doi [Phys. Rev. D] {10.1103/PhysRevD.100.123522}, 100, 123522

\bibitem[\protect\citeauthoryear{Berti, Spinelli, Haridasu, Viel  \&
  Silvestri}{Berti et~al.}{2022}]{Berti:2021}
Berti M.,  Spinelli M.,  Haridasu B.~S.,  Viel M.,   Silvestri A.,  2022,
  \mn@doi [JCAP] {10.1088/1475-7516/2022/01/018}, 01, 018

\bibitem[\protect\citeauthoryear{Berti, Spinelli  \& Viel}{Berti
  et~al.}{2023}]{Berti:2022ilk}
Berti M.,  Spinelli M.,   Viel M.,  2023, \mn@doi [Mon. Not. Roy. Astron. Soc.]
  {10.1093/mnras/stad685}, 521, 3221

\bibitem[\protect\citeauthoryear{Bharadwaj, Nath, Nath  \& Sethi}{Bharadwaj
  et~al.}{2001}]{Bharadwaj:2000}
Bharadwaj S.,  Nath B.~B.,  Nath B.~B.,   Sethi S.~K.,  2001, \mn@doi [J.
  Astrophys. Astron.] {10.1007/BF02933588}, 22, 21

\bibitem[\protect\citeauthoryear{Blake}{Blake}{2019}]{Blake:2019}
Blake C.,  2019, \mn@doi [Mon. Not. Roy. Astron. Soc.] {10.1093/mnras/stz2145},
  489, 153

\bibitem[\protect\citeauthoryear{Blake et~al.}{Blake
  et~al.}{2011}]{Blake:2011rj}
Blake C.,  et~al., 2011, \mn@doi [Mon. Not. Roy. Astron. Soc.]
  {10.1111/j.1365-2966.2011.18903.x}, 415, 2876

\bibitem[\protect\citeauthoryear{Blanchard et~al.}{Blanchard
  et~al.}{2020}]{Euclid:2019clj}
Blanchard A.,  et~al., 2020, \mn@doi [Astron. Astrophys.]
  {10.1051/0004-6361/202038071}, 642, A191

\bibitem[\protect\citeauthoryear{Bull et~al.}{Bull et~al.}{2016}]{bull:2015}
Bull P.,  et~al., 2016, \mn@doi [Phys. Dark Univ.]
  {10.1016/j.dark.2016.02.001}, 12, 56

\bibitem[\protect\citeauthoryear{Carucci, Irfan  \& Bobin}{Carucci
  et~al.}{2020}]{Carucci2020}
Carucci I.~P.,  Irfan M.~O.,   Bobin J.,  2020, \mn@doi [Mon. Not. Roy. Astron.
  Soc.] {10.1093/mnras/staa2854}, 499, 304

\bibitem[\protect\citeauthoryear{Casas, Carucci, Pettorino, Camera  \&
  Martinelli}{Casas et~al.}{2023}]{Casas:2022vik}
Casas S.,  Carucci I.~P.,  Pettorino V.,  Camera S.,   Martinelli M.,  2023,
  \mn@doi [Phys. Dark Univ.] {10.1016/j.dark.2022.101151}, 39, 101151

\bibitem[\protect\citeauthoryear{Castorina \& Villaescusa-Navarro}{Castorina \&
  Villaescusa-Navarro}{2017}]{Castorina:2016bfm}
Castorina E.,  Villaescusa-Navarro F.,  2017, \mn@doi [Mon. Not. Roy. Astron.
  Soc.] {10.1093/mnras/stx1599}, 471, 1788

\bibitem[\protect\citeauthoryear{Chang, Pen, Peterson  \& McDonald}{Chang
  et~al.}{2008}]{Chang:2007}
Chang T.-C.,  Pen U.-L.,  Peterson J.~B.,   McDonald P.,  2008, \mn@doi [Phys.
  Rev. Lett.] {10.1103/PhysRevLett.100.091303}, 100, 091303

\bibitem[\protect\citeauthoryear{{Chang}, {Pen}, {Bandura}  \&
  {Peterson}}{{Chang} et~al.}{2010}]{Chang2010}
{Chang} T.-C.,  {Pen} U.-L.,  {Bandura} K.,   {Peterson} J.~B.,  2010, \mn@doi
  [\nat] {10.1038/nature09187}, \href
  {https://ui.adsabs.harvard.edu/abs/2010Natur.466..463C} {466, 463}

\bibitem[\protect\citeauthoryear{{Cunnington}}{{Cunnington}}{2022}]{Cunnington:2022}
{Cunnington} S.,  2022, \mn@doi [\mnras] {10.1093/mnras/stac576}, \href
  {https://ui.adsabs.harvard.edu/abs/2022MNRAS.512.2408C} {512, 2408}

\bibitem[\protect\citeauthoryear{Cunnington, Pourtsidou, Soares, Blake  \&
  Bacon}{Cunnington et~al.}{2020}]{Cunnington:2020}
Cunnington S.,  Pourtsidou A.,  Soares P.~S.,  Blake C.,   Bacon D.,  2020,
  \mn@doi [Mon. Not. Roy. Astron. Soc.] {10.1093/mnras/staa1524}, 496, 415

\bibitem[\protect\citeauthoryear{Cunnington, Irfan, Carucci, Pourtsidou  \&
  Bobin}{Cunnington et~al.}{2021}]{Cunnington2021}
Cunnington S.,  Irfan M.~O.,  Carucci I.~P.,  Pourtsidou A.,   Bobin J.,  2021,
  \mn@doi [Mon. Not. Roy. Astron. Soc.] {10.1093/mnras/stab856}, 504, 208

\bibitem[\protect\citeauthoryear{Cunnington et~al.}{Cunnington
  et~al.}{2022}]{Cunnington:2022uzo}
Cunnington S.,  et~al., 2022, \mn@doi [Mon. Not. Roy. Astron. Soc.]
  {10.1093/mnras/stac3060}, 518, 6262

\bibitem[\protect\citeauthoryear{D'Amico, Gleyzes, Kokron, Markovic, Senatore,
  Zhang, Beutler  \& Gil-Mar\'\i{}n}{D'Amico et~al.}{2020}]{DAmico:2019fhj}
D'Amico G.,  Gleyzes J.,  Kokron N.,  Markovic K.,  Senatore L.,  Zhang P.,
  Beutler F.,   Gil-Mar\'\i{}n H.,  2020, \mn@doi [JCAP]
  {10.1088/1475-7516/2020/05/005}, 05, 005

\bibitem[\protect\citeauthoryear{{Drinkwater} et~al.,}{{Drinkwater}
  et~al.}{2010}]{Drinkwater2010:wigglez}
{Drinkwater} M.~J.,  et~al., 2010, \mn@doi [\mnras]
  {10.1111/j.1365-2966.2009.15754.x}, \href
  {https://ui.adsabs.harvard.edu/abs/2010MNRAS.401.1429D} {401, 1429}

\bibitem[\protect\citeauthoryear{Drinkwater et~al.}{Drinkwater
  et~al.}{2018}]{WiggleZ:2018def}
Drinkwater M.~J.,  et~al., 2018, \mn@doi [Mon. Not. Roy. Astron. Soc.]
  {10.1093/mnras/stx2963}, 474, 4151

\bibitem[\protect\citeauthoryear{Furlanetto, Oh  \& Briggs}{Furlanetto
  et~al.}{2006}]{Furlanetto:2006}
Furlanetto S.,  Oh S.~P.,   Briggs F.,  2006, \mn@doi [Phys. Rept.]
  {10.1016/j.physrep.2006.08.002}, 433, 181

\bibitem[\protect\citeauthoryear{Gil-Mar\'\i{}n, Percival, Verde, Brownstein,
  Chuang, Kitaura, Rodr\'\i{}guez-Torres  \& Olmstead}{Gil-Mar\'\i{}n
  et~al.}{2017}]{Gil-Marin:2016wya}
Gil-Mar\'\i{}n H.,  Percival W.~J.,  Verde L.,  Brownstein J.~R.,  Chuang
  C.-H.,  Kitaura F.-S.,  Rodr\'\i{}guez-Torres S.~A.,   Olmstead M.~D.,  2017,
  \mn@doi [Mon. Not. Roy. Astron. Soc.] {10.1093/mnras/stw2679}, 465, 1757

\bibitem[\protect\citeauthoryear{Gilks, Richardson  \& Spiegelhalter}{Gilks
  et~al.}{1995}]{gilks:1995}
Gilks W.,  Richardson S.,   Spiegelhalter D.,  1995, Markov Chain Monte Carlo
  in Practice.
Chapman \& Hall/CRC Interdisciplinary Statistics, Taylor \& Francis, \url
  {https://books.google.it/books?id=TRXrMWY\_i2IC}

\bibitem[\protect\citeauthoryear{Hand, Seljak, Beutler  \& Vlah}{Hand
  et~al.}{2017}]{Hand:2017ilm}
Hand N.,  Seljak U.,  Beutler F.,   Vlah Z.,  2017, \mn@doi [JCAP]
  {10.1088/1475-7516/2017/10/009}, 10, 009

\bibitem[\protect\citeauthoryear{Hu, Wang, Wu, Wang, Zhang  \& Chen}{Hu
  et~al.}{2020}]{Hu:2019okh}
Hu W.,  Wang X.,  Wu F.,  Wang Y.,  Zhang P.,   Chen X.,  2020, \mn@doi [Mon.
  Not. Roy. Astron. Soc.] {10.1093/mnras/staa650}, 493, 5854

\bibitem[\protect\citeauthoryear{Irfan \& Bull}{Irfan \&
  Bull}{2021}]{Irfan2021}
Irfan M.~O.,  Bull P.,  2021, \mn@doi [Mon. Not. Roy. Astron. Soc.]
  {10.1093/mnras/stab2855}, 508, 3551

\bibitem[\protect\citeauthoryear{{Irfan} et~al.,}{{Irfan}
  et~al.}{2022}]{irfan2022}
{Irfan} M.~O.,  et~al., 2022, \mn@doi [\mnras] {10.1093/mnras/stab3346}, \href
  {https://ui.adsabs.harvard.edu/abs/2022MNRAS.509.4923I} {509, 4923}

\bibitem[\protect\citeauthoryear{Jiang et~al.}{Jiang
  et~al.}{2023}]{Jiang:2023zex}
Jiang Y.-E.,  et~al., 2023, \mn@doi [Res. Astron. Astrophys.]
  {10.1088/1674-4527/accdc0}, 23, 075003

\bibitem[\protect\citeauthoryear{Jolicoeur, Maartens  \& Dlamini}{Jolicoeur
  et~al.}{2023}]{Jolicoeur:2023tcu}
Jolicoeur S.,  Maartens R.,   Dlamini S.,  2023, \mn@doi [Eur. Phys. J. C]
  {10.1140/epjc/s10052-023-11482-2}, 83, 320

\bibitem[\protect\citeauthoryear{Kaiser}{Kaiser}{1987}]{kaiser1987}
Kaiser N.,  1987, Mon. Not. Roy. Astron. Soc., 227, 1

\bibitem[\protect\citeauthoryear{Karagiannis, Slosar  \& Liguori}{Karagiannis
  et~al.}{2020}]{Karagiannis:2019jjx}
Karagiannis D.,  Slosar A.,   Liguori M.,  2020, \mn@doi [JCAP]
  {10.1088/1475-7516/2020/11/052}, 11, 052

\bibitem[\protect\citeauthoryear{Karagiannis, Maartens  \&
  Randrianjanahary}{Karagiannis et~al.}{2022}]{Karagiannis:2022ylq}
Karagiannis D.,  Maartens R.,   Randrianjanahary L.~F.,  2022, \mn@doi [JCAP]
  {10.1088/1475-7516/2022/11/003}, 11, 003

\bibitem[\protect\citeauthoryear{Khandai, Sethi, Di~Matteo, Croft, Springel,
  Jana  \& Gardner}{Khandai et~al.}{2011}]{Khandai:2010hs}
Khandai N.,  Sethi S.~K.,  Di~Matteo T.,  Croft R. A.~C.,  Springel V.,  Jana
  A.,   Gardner J.~P.,  2011, \mn@doi [Mon. Not. Roy. Astron. Soc.]
  {10.1111/j.1365-2966.2011.18881.x}, 415, 2580

\bibitem[\protect\citeauthoryear{Kovetz et~al.}{Kovetz
  et~al.}{2017}]{Kovetz:2017}
Kovetz E.~D.,  et~al., 2017, {Line-Intensity Mapping: 2017 Status Report}
  (\mn@eprint {arXiv} {1709.09066})

\bibitem[\protect\citeauthoryear{Lewis}{Lewis}{2013}]{Lewis:2013}
Lewis A.,  2013, \mn@doi [Phys. Rev. D] {10.1103/PhysRevD.87.103529}, 87,
  103529

\bibitem[\protect\citeauthoryear{Lewis \& Bridle}{Lewis \&
  Bridle}{2002}]{Lewis:2002}
Lewis A.,  Bridle S.,  2002, \mn@doi [Phys. Rev. D]
  {10.1103/PhysRevD.66.103511}, 66, 103511

\bibitem[\protect\citeauthoryear{Lewis, Challinor  \& Lasenby}{Lewis
  et~al.}{2000}]{lewis:2000}
Lewis A.,  Challinor A.,   Lasenby A.,  2000, \mn@doi [Astrophys. J.]
  {10.1086/309179}, 538, 473

\bibitem[\protect\citeauthoryear{{Masui} et~al.,}{{Masui}
  et~al.}{2013}]{Masui2013}
{Masui} K.~W.,  et~al., 2013, \mn@doi [\apjl] {10.1088/2041-8205/763/1/L20},
  \href {https://ui.adsabs.harvard.edu/abs/2013ApJ...763L..20M} {763, L20}

\bibitem[\protect\citeauthoryear{Matshawule, Spinelli, Santos  \&
  Ngobese}{Matshawule et~al.}{2020}]{Matshawule:2021}
Matshawule S.~D.,  Spinelli M.,  Santos M.~G.,   Ngobese S.,  2020, \mn@doi
  [Mon. Not. Roy. Astron. Soc.] {10.1093/mnras/stab1688}

\bibitem[\protect\citeauthoryear{McQuinn, Zahn, Zaldarriaga, Hernquist  \&
  Furlanetto}{McQuinn et~al.}{2006}]{McQuinn:2005}
McQuinn M.,  Zahn O.,  Zaldarriaga M.,  Hernquist L.,   Furlanetto S.~R.,
  2006, \mn@doi [Astrophys. J.] {10.1086/505167}, 653, 815

\bibitem[\protect\citeauthoryear{Mead, Heymans, Lombriser, Peacock, Steele  \&
  Winther}{Mead et~al.}{2016}]{Mead:2016}
Mead A.,  Heymans C.,  Lombriser L.,  Peacock J.,  Steele O.,   Winther H.,
  2016, \mn@doi [Mon. Not. Roy. Astron. Soc.] {10.1093/mnras/stw681}, 459, 1468

\bibitem[\protect\citeauthoryear{Newburgh et~al.}{Newburgh
  et~al.}{2016}]{Newburgh:2016mwi}
Newburgh L.~B.,  et~al., 2016, \mn@doi [Proc. SPIE Int. Soc. Opt. Eng.]
  {10.1117/12.2234286}, 9906, 99065X

\bibitem[\protect\citeauthoryear{{Obuljen}, {Castorina}, {Villaescusa-Navarro}
  \& {Viel}}{{Obuljen} et~al.}{2018}]{obuljen18}
{Obuljen} A.,  {Castorina} E.,  {Villaescusa-Navarro} F.,   {Viel} M.,  2018,
  \mn@doi [\jcap] {10.1088/1475-7516/2018/05/004}, \href
  {https://ui.adsabs.harvard.edu/abs/2018JCAP...05..004O} {2018, 004}

\bibitem[\protect\citeauthoryear{Padmanabhan, Maartens, Umeh  \&
  Camera}{Padmanabhan et~al.}{2023}]{Padmanabhan:2023hfr}
Padmanabhan H.,  Maartens R.,  Umeh O.,   Camera S.,  2023, {The HI intensity
  mapping power spectrum: insights from recent measurements} (\mn@eprint
  {arXiv} {2305.09720})

\bibitem[\protect\citeauthoryear{{Paul}, {Santos}, {Chen}  \& {Wolz}}{{Paul}
  et~al.}{2023}]{paul23}
{Paul} S.,  {Santos} M.~G.,  {Chen} Z.,   {Wolz} L.,  2023, \mn@doi [arXiv
  e-prints] {10.48550/arXiv.2301.11943}, \href
  {https://ui.adsabs.harvard.edu/abs/2023arXiv230111943P} {p. arXiv:2301.11943}

\bibitem[\protect\citeauthoryear{Planck Collaboration~III}{Planck
  Collaboration~III}{2020}]{planck:2018_maps}
Planck Collaboration~III .,  2020, \mn@doi [Astron. Astrophys.]
  {10.1051/0004-6361/201832909}, 641, A3

\bibitem[\protect\citeauthoryear{Planck Collaboration~V}{Planck
  Collaboration~V}{2020}]{planck:2018like}
Planck Collaboration~V .,  2020, \mn@doi [Astron. Astrophys.]
  {10.1051/0004-6361/201936386}, 641, A5

\bibitem[\protect\citeauthoryear{Planck Collaboration~VI}{Planck
  Collaboration~VI}{2020}]{planck:2018}
Planck Collaboration~VI .,  2020, \mn@doi [\aap] {10.1051/0004-6361/201833910},
  \href {https://ui.adsabs.harvard.edu/abs/2020A&A...641A...6P} {641, A6}

\bibitem[\protect\citeauthoryear{{Pritchard} \& {Loeb}}{{Pritchard} \&
  {Loeb}}{2012}]{review}
{Pritchard} J.~R.,  {Loeb} A.,  2012, \mn@doi [Reports on Progress in Physics]
  {10.1088/0034-4885/75/8/086901}, \href
  {https://ui.adsabs.harvard.edu/abs/2012RPPh...75h6901P} {75, 086901}

\bibitem[\protect\citeauthoryear{{Raichoor} et~al.,}{{Raichoor}
  et~al.}{2023}]{desiemission}
{Raichoor} A.,  et~al., 2023, \mn@doi [\aj] {10.3847/1538-3881/acb213}, \href
  {https://ui.adsabs.harvard.edu/abs/2023AJ....165..126R} {165, 126}

\bibitem[\protect\citeauthoryear{SKA Cosmology~SWG}{SKA
  Cosmology~SWG}{2020}]{Bacon:2018}
SKA Cosmology~SWG .,  2020, \mn@doi [Publ. Astron. Soc. Austral.]
  {10.1017/pasa.2019.51}, 37, e007

\bibitem[\protect\citeauthoryear{Santos et~al.}{Santos
  et~al.}{2015}]{Santos:2015}
Santos M.~G.,  et~al., 2015, \mn@doi [PoS] {10.22323/1.215.0019}, AASKA14, 019

\bibitem[\protect\citeauthoryear{Santos et~al.}{Santos
  et~al.}{2017}]{Santos:2017}
Santos M.~G.,  et~al., 2017, in {MeerKAT Science}: {On the Pathway to the SKA}.
   (\mn@eprint {arXiv} {1709.06099})

\bibitem[\protect\citeauthoryear{Seo, Dodelson, Marriner, Mcginnis, Stebbins,
  Stoughton  \& Vallinotto}{Seo et~al.}{2010}]{Seo:2009fq}
Seo H.-J.,  Dodelson S.,  Marriner J.,  Mcginnis D.,  Stebbins A.,  Stoughton
  C.,   Vallinotto A.,  2010, \mn@doi [Astrophys. J.]
  {10.1088/0004-637X/721/1/164}, 721, 164

\bibitem[\protect\citeauthoryear{{Smith}}{{Smith}}{2009}]{Smith09}
{Smith} R.~E.,  2009, \mn@doi [\mnras] {10.1111/j.1365-2966.2009.15490.x},
  \href {https://ui.adsabs.harvard.edu/abs/2009MNRAS.400..851S} {400, 851}

\bibitem[\protect\citeauthoryear{Smith et~al.,}{Smith
  et~al.}{2003}]{Smith:2002}
Smith R.~E.,  et~al., 2003, \mn@doi [Mon. Not. Roy. Astron. Soc.]
  {10.1046/j.1365-8711.2003.06503.x}, 341, 1311

\bibitem[\protect\citeauthoryear{Soares, Cunnington, Pourtsidou  \&
  Blake}{Soares et~al.}{2021}]{Soares:2020}
Soares P.~S.,  Cunnington S.,  Pourtsidou A.,   Blake C.,  2021, \mn@doi [Mon.
  Not. Roy. Astron. Soc.] {10.1093/mnras/stab027}, 502, 2549

\bibitem[\protect\citeauthoryear{Soares, Watkinson, Cunnington  \&
  Pourtsidou}{Soares et~al.}{2022}]{Soares2021GPR}
Soares P.~S.,  Watkinson C.~A.,  Cunnington S.,   Pourtsidou A.,  2022, \mn@doi
  [Mon. Not. Roy. Astron. Soc.] {10.1093/mnras/stab2594}, 510, 5872

\bibitem[\protect\citeauthoryear{Spinelli, Zoldan, De~Lucia, Xie  \&
  Viel}{Spinelli et~al.}{2020}]{Spinelli:2019}
Spinelli M.,  Zoldan A.,  De~Lucia G.,  Xie L.,   Viel M.,  2020, \mn@doi [Mon.
  Not. Roy. Astron. Soc.] {10.1093/mnras/staa604}, 493, 5434

\bibitem[\protect\citeauthoryear{Spinelli, Carucci, Cunnington, Harper, Irfan,
  Fonseca, Pourtsidou  \& Wolz}{Spinelli et~al.}{2021}]{Spinelli:2021emp}
Spinelli M.,  Carucci I.~P.,  Cunnington S.,  Harper S.~E.,  Irfan M.~O.,
  Fonseca J.,  Pourtsidou A.,   Wolz L.,  2021, \mn@doi [Mon. Not. Roy. Astron.
  Soc.] {10.1093/mnras/stab3064}, 509, 2048

\bibitem[\protect\citeauthoryear{Vargas-Maga\~na, Brooks, Levi  \&
  Tarle}{Vargas-Maga\~na et~al.}{2018}]{Vargas-Magana:2018rbb}
Vargas-Maga\~na M.,  Brooks D.~D.,  Levi M.~M.,   Tarle G.~G.,  2018, in {53rd
  Rencontres de Moriond on Cosmology}. pp 11--18 (\mn@eprint {arXiv}
  {1901.01581})

\bibitem[\protect\citeauthoryear{Viljoen, Fonseca  \& Maartens}{Viljoen
  et~al.}{2020}]{Viljoen:2020efi}
Viljoen J.-A.,  Fonseca J.,   Maartens R.,  2020, \mn@doi [JCAP]
  {10.1088/1475-7516/2020/09/054}, 09, 054

\bibitem[\protect\citeauthoryear{{Villaescusa-Navarro}, {Viel}, {Alonso},
  {Datta}, {Bull}  \& {Santos}}{{Villaescusa-Navarro} et~al.}{2015}]{villa15}
{Villaescusa-Navarro} F.,  {Viel} M.,  {Alonso} D.,  {Datta} K.~K.,  {Bull} P.,
    {Santos} M.~G.,  2015, \mn@doi [\jcap] {10.1088/1475-7516/2015/03/034},
  \href {https://ui.adsabs.harvard.edu/abs/2015JCAP...03..034V} {2015, 034}

\bibitem[\protect\citeauthoryear{Villaescusa-Navarro, Alonso  \&
  Viel}{Villaescusa-Navarro et~al.}{2017}]{Villaescusa-Navarro:2016}
Villaescusa-Navarro F.,  Alonso D.,   Viel M.,  2017, \mn@doi [Mon. Not. Roy.
  Astron. Soc.] {10.1093/mnras/stw3224}, 466, 2736

\bibitem[\protect\citeauthoryear{Villaescusa-Navarro
  et~al.}{Villaescusa-Navarro et~al.}{2018}]{Villaescusa-Navarro:2018}
Villaescusa-Navarro F.,  et~al., 2018, \mn@doi [Astrophys. J.]
  {10.3847/1538-4357/aadba0}, 866, 135

\bibitem[\protect\citeauthoryear{Wang et~al.}{Wang et~al.}{2021}]{Wang:2021}
Wang J.,  et~al., 2021, \mn@doi [Mon. Not. Roy. Astron. Soc.]
  {10.1093/mnras/stab1365}, 505, 3698

\bibitem[\protect\citeauthoryear{Wolz, Tonini, Blake  \& Wyithe}{Wolz
  et~al.}{2016}]{Wolz2016}
Wolz L.,  Tonini C.,  Blake C.,   Wyithe J. S.~B.,  2016, \mn@doi [Mon. Not.
  Roy. Astron. Soc.] {10.1093/mnras/stw535}, 458, 3399

\bibitem[\protect\citeauthoryear{{Wolz} et~al.,}{{Wolz}
  et~al.}{2022}]{Wolz2022}
{Wolz} L.,  et~al., 2022, \mn@doi [\mnras] {10.1093/mnras/stab3621}, \href
  {https://ui.adsabs.harvard.edu/abs/2022MNRAS.510.3495W} {510, 3495}

\bibitem[\protect\citeauthoryear{Wu, Li, Zhang  \& Zhang}{Wu
  et~al.}{2023}]{Wu:2022jkf}
Wu P.-J.,  Li Y.,  Zhang J.-F.,   Zhang X.,  2023, \mn@doi [Sci. China Phys.
  Mech. Astron.] {10.1007/s11433-022-2104-7}, 66, 270413

\makeatother
\end{thebibliography}

%%%%%%%%%%%%%%%%%%%%%%%%%%%%%%%%%%%%%%%%%%%%%%%%%%

%%%%%%%%%%%%%%%%% APPENDICES %%%%%%%%%%%%%%%%%%%%%

\appendix
\section{On the impact of including the Alcock-Paczynski effects}
\label{sec:appendix_AP}
\begin{figure}
	\includegraphics[width=\columnwidth]{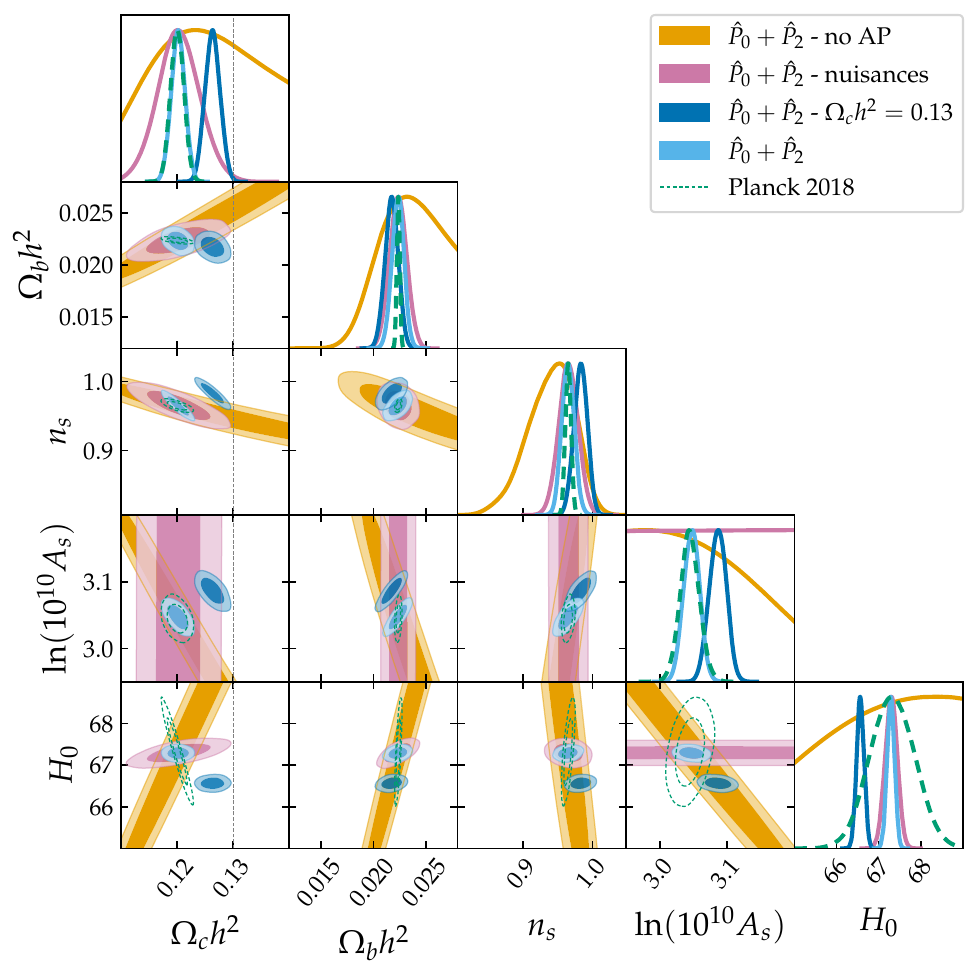}
 \includegraphics[width=\columnwidth]{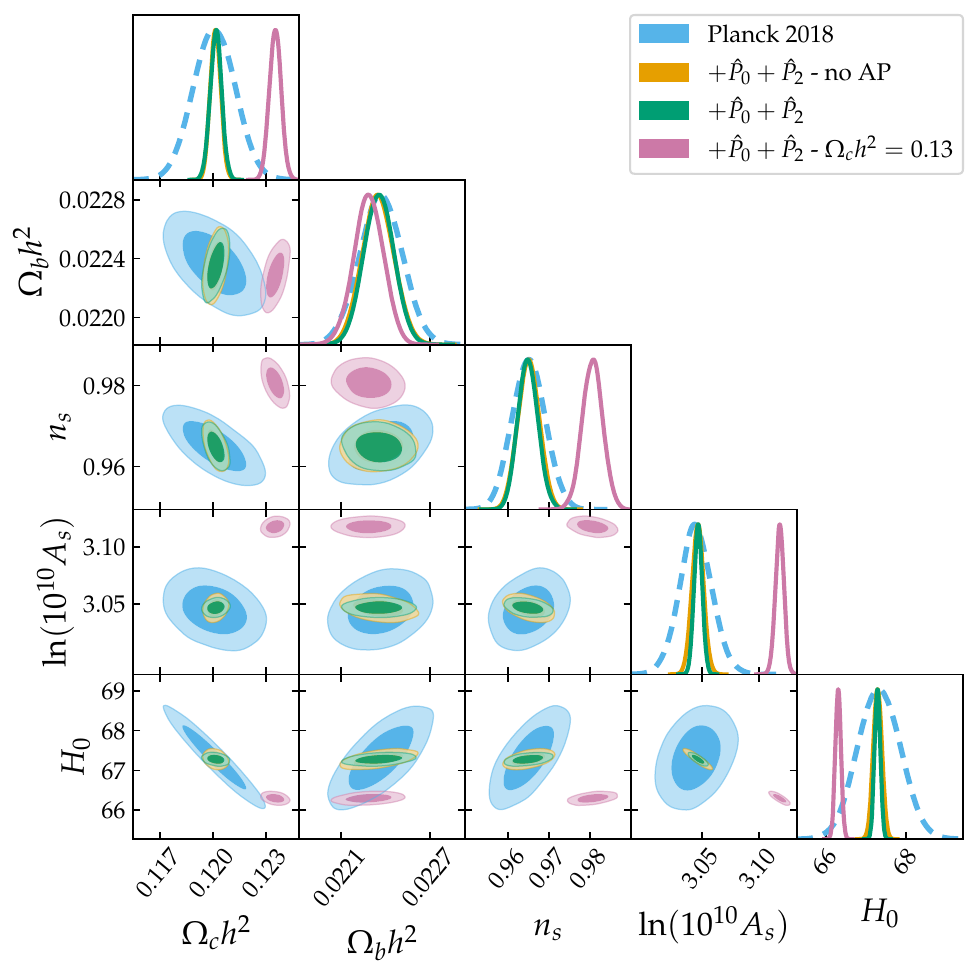}
    \caption{Joint constraints (68\% and 95\% confidence regions) and marginalised posterior distributions on a subset of the cosmological parameters. We show results for the 21cm multipoles alone (upper panel) and combined with {CMB} observations (lower panel). The label "Planck 2018" (dashed lines) stands for TT, TE, EE + lowE + lensing, while the label "$\Hat{P}_0 + \Hat{P}_2$"  stands for the forecasted 21cm non-linear power spectrum monopole and quadrupole observations, with and without ("no AP") {AP} effects taken into account.  "$\Hat{P}_0 + \Hat{P}_2 - \Omega_c h^2 = 0.13$" labels the results obtained for a mock data set with a value of $\Omega_c h^2$ different from the assumed fiducial cosmology. The label "nuis." indicates that we vary the nuisance parameters along with the cosmological ones.}
    \label{fig:app_AP}
\end{figure}
With respect to the analysis in \cite{Berti:2022ilk}, in this work, we extended our likelihood code to include the Alcock-Paczynski distortions, which are used in other works~(e.g. \citep{Bernal:2019,Soares:2020,Casas:2022vik}). We neglected {AP} effects in the first approximation because we assumed to know the true cosmology, given that it is the one we input when constructing the mock data set. In this section, we give an overview of how the cosmological parameter constraints are affected by the addition of {AP} effects. 

Contrary to what we naively expected, implementing the {AP} modifications significantly improves the constraints. In the upper panel of \autoref{fig:app_AP}, we compare the effects of different mock data sets. Our reference (orange lines and contours) is the 21cm power spectrum monopole ($\hat{P}_0$) and quadrupole ($\hat{P}_2$) mock data set that we construct in \cite{Berti:2022ilk}. This data set forecasts {SKAO} observations in multipole redshift bins in the range 0 - 3, i.e. for six bins centred at $\{0.25,\, 0.75,\, 1.25,\, 1.75,\, 2.25,\, 2.75 \}$ and with a width of $\Delta z = 0.5$. The nuisances parameters for this data set are $\Bar{T}_b b_{\rm{HI}} \sigma_8$, $\Bar{T}_b f \sigma_8$, and the {HI} shot noise, for the non-linear 21cm power spectrum. 

Using the exact same framework, but implementing also the {AP} distortions of the amplitude and of the wave vectors as described in \secref{sec:P21_model_cross} (light blue contours), we find a crucially improved constraining power. E.g., for $\Omega_ch^2$ with only 21cm observations, we recover the Planck constraint (dashed green lines and contours). On $H_0$, instead, we find even better constraints than Planck. When adding the nuisances (pink contours), the improvement is reduced, although still significantly better than the no {AP} case. We believe that the extra dependence on $H(z)$ that is introduced in the observable with the {AP} modifications is the cause of the improved constraining power.

Dealing with mock observations fabricated by ourselves, we have the advantage of knowing the true cosmology. We, thus, further test the impact of {AP} by creating a data set with a different value of $\Omega_c h^2 = 0.13$. We, however, do not change the fiducial cosmology, for which  $\Omega_c h^2 = 0.12011$. Running the {MCMC} analysis on this data set we find consistent results. The $\Omega_c h^2$ constraint is pushed towards the true value, resulting in a constraint in between the true value and the assumed fiducial one. The errors, instead, are left unchanged, although the 2D contours are rigidly shifted. Thus, it seems that assuming the wrong cosmology impacts only the mean marginalised values of the parameters. 

However, the smoking gun of having assumed the wrong cosmology is the probability distribution of the {AP} parameters themselves. Although we do not show the results here, we implemented the time-dependent $\alpha_\perp$ and $\alpha_\parallel$ as derived parameters and computed the marginalised constraints. We find that when the true cosmology matches the fiducial one, the $\alpha_\perp$, and $\alpha_\parallel$ marginalised posteriors are centred around one. When, instead, the true cosmology is not the assumed one, although still compatible with one the constraints, both 1D and 2D, are clearly shifted. Thus, one can test their assumptions by looking at the {AP} parameters constraints.  

The lower panel of \autoref{fig:app_AP} shows the results for the same exercise, but combining 21cm observations with Planck data.

We conclude that even when the whole set of cosmological parameters is used, the {AP} distortions are crucial not only to take into account our lack of knowledge of the true underlying cosmology but also to increase the constraining power on the cosmological parameters in matter power spectrum dependent probes as the 21cm {IM} one. 

\section{Test on the mock data construction procedure}
\label{app:test}
\begin{figure}
	\includegraphics[width=\columnwidth]{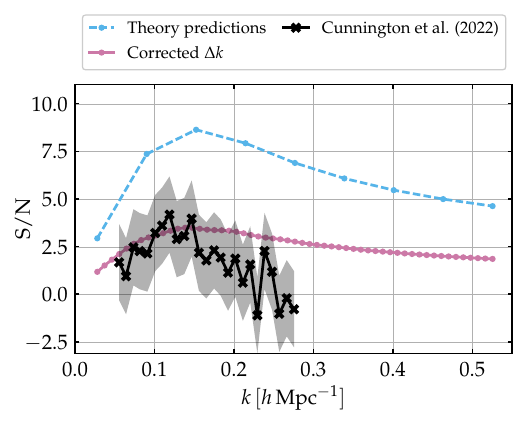}
    \caption{Signal-to-noise ratio as a function of $k$. We compare real data observations ("Cunnington et al. (2022)"), with the signal-to-noise predicted by the formalism adopted in this work. The gray shaded area shows the $2\sigma$ region for the observed signal-to-noise.}
    \label{fig:signal_to_noise_real_data}
	\includegraphics[width=\columnwidth]{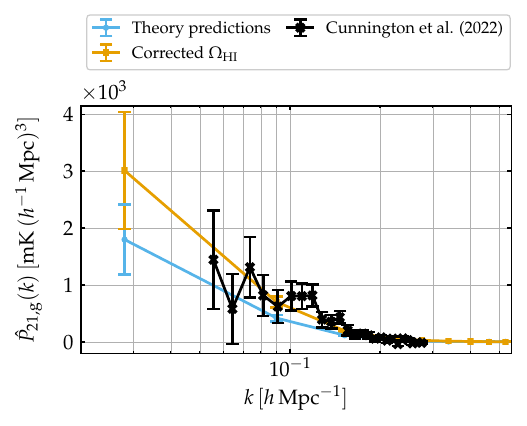}
    \caption{Observed and predicted 21cm MeerKAT observations in cross-correlation with WiggleZ galaxy clustering. }
    \label{fig:real_data_vs_mock}
\end{figure}
To test the procedure we follow to construct the mock data sets, we compare our predictions to the measured cross-correlation data published in~\cite{Cunnington:2022uzo}. 

As in \secref{sec:res_real_data_constraints}, we adjust the parameters of our formalism to mimic MeerKAT observations in the redshift bin centred at $z=0.43$ and with $\Delta z = 0.059$. We find that with our pipeline we predict fewer $k$-bin in a wider scale range and a slightly different value for the brightness temperature, due to { a small difference in the theory-predicted and measured value of $\Omega_{\rm{HI}}$}. Correcting for the brightness temperature results in cross-correlation power spectrum values more in agreement with observations (\autoref{fig:real_data_vs_mock}). However, this is not enough to reproduce the observed signal-to-noise ratio. Indeed, we find that varying the brightness temperature changes the power spectrum and the errors consistently, not impacting the signal-to-noise. Instead, adjusting the $k$-bin width to match the one in~\cite{Cunnington:2022uzo} is enough to better reproduce the observed signal-to-noise ratio, as shown in \autoref{fig:signal_to_noise_real_data}.

We conclude that, compared to the state of the art, the pipeline we adopt in this work is consistent with real observational data. We are, however, more optimistic about the accessible scales and bin widths. Differences in the predicted power spectrum amplitude, i.e. the brightness temperature, are taken into account when opening the parameter space to the nuisance parameters.

\section{On the impact of foreground contamination}
\begin{figure}
	\includegraphics[width=\columnwidth]{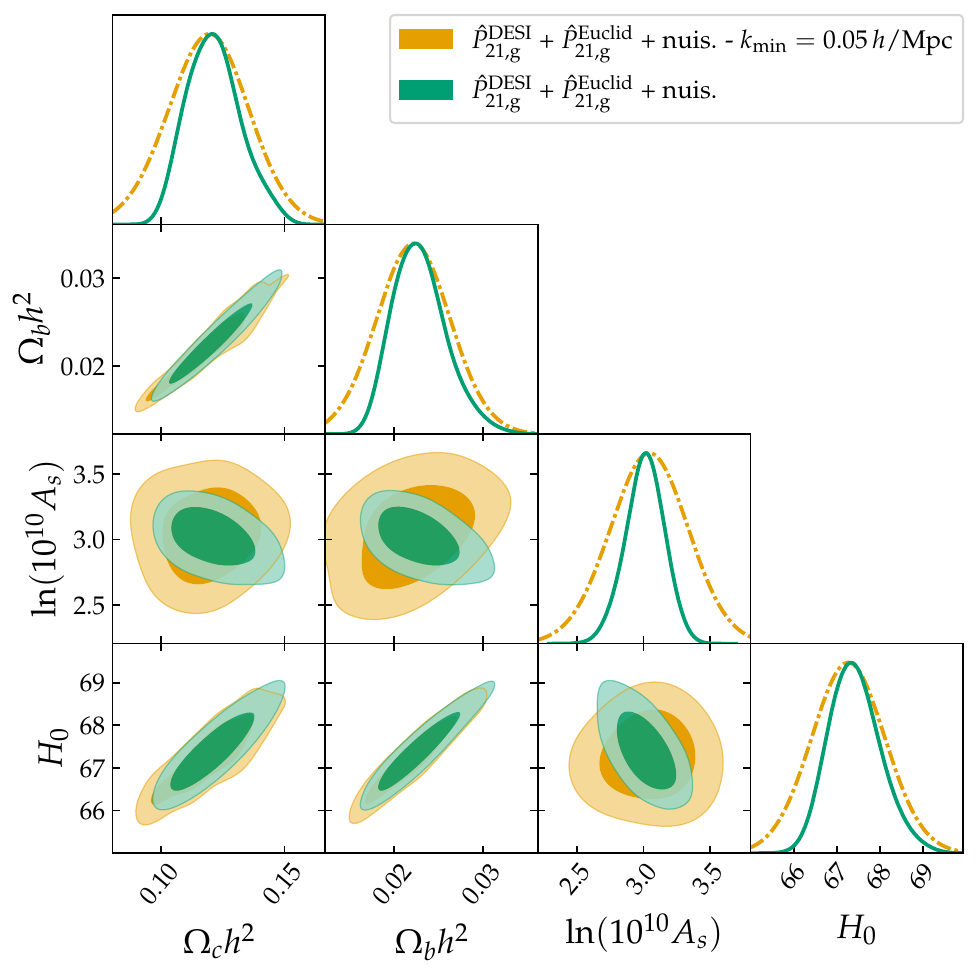}
    \caption{{Joint constraints (68\% and 95\% confidence regions) and marginalised posterior distributions on a subset of the cosmological parameters. We compare the results presented in \autoref{sec:res_cross} (green, continuous lines) with what is obtained mimicking the effect of the foregrounds by cutting the data set at $k_{\rm min} = 0.05 \, h/$Mpc (yellow, dashed lines). The label "nuis." indicates that we vary the nuisance parameters along with the cosmological ones.}}
    \label{fig:cuts}
\end{figure}
The aim of the work presented in this paper is the development and testing of a pipeline for future 21cm and galaxy clustering cross-correlation assuming the ideal scenario where all the systematics in the data have been already treated. It is however interesting to explore the effect of the key challenge of intensity mapping i.e. foreground contamination.  

We follow e.g. \cite{Karagiannis:2019jjx} and mimic the effect of the foreground as a loss of observed large scales by cutting our mock dataset at $k_{\rm min} = 0.05 \, h/$Mpc. \autoref{fig:cuts} shows the comparison between the result presented above and the scale-cut data set for a reference case. As expected, foreground removal worsens the constraints on the cosmological parameter. However, it is worth noticing that the correlation between $\Omega_c h^2$ and $H_0$ is left unchanged.

As a consequence, when combining the scale-cut cross-correlation mock data set with measured Planck data, we observe that the same results are achieved with or without accounting for foreground effects. We can then conclude that, although real data could be affected by more complex systematics, the loss of some of the large scale in the cross-power spectrum does not prevent the increase in constraining power obtained combined with Planck results.

{

\section{Comparison with galaxy clustering}

In \secref{sec:res_cross}, we compare results on the cosmological parameter constraints from the cross-correlation of 21cm and galaxy clustering observations with the ones obtained from 21cm auto-power spectrum measurements. For completeness, in what follows we also forecast the constraining power of galaxy clustering auto-power spectrum measurements. 

The pipeline we built for the 21cm observables can be extended to include the analysis of galaxy clustering auto-power spectrum monopole $P_{\rm g} (k,z)$ by implementing an additional likelihood function. For this purpose, we construct a toy data set of future DESI and Euclid-like observations keeping the same framework of observed redshifts, scales and volumes presented in \secref{sec:methods}. We highlight that a more realistic forecast that includes the full constraining power of $P_{\rm g} (k,z)$ measurements goes beyond the goal of this exercise.

As in \secref{sec:methods}, our galaxy clustering data set is constructed from a theory predictions for the galaxy clustering auto-power spectrum $\hat{P}_{\rm g}(k,z)$ (\autoref{eq:P_galaxy}) and using the survey specifications for values of the bias and the shot noise (\autoref{sec:surveys}). Following the formalism used for the 21cm auto-power spectrum \citep{Berti:2021}, errors on each single data point are estimated as $$ \hat{\sigma}_{\rm g}(k) = \frac{1}{\sqrt{2N_{\rm modes}(k)}} \sqrt{  \int_{-1}^1  {\rm d}\mu\, \left(\hat{P}_{\rm g}(k) + \frac{1}{\bar{n}_{\rm g}}\right)^2}.$$
We stress that to be consistent with our settings, we consider limited sky volumes and do not investigate fully non-linear scales. We thus expect to obtain with our mock data set pessimistic constraints with respect to the full potential of stage IV surveys.

Our results for the mock Euclid and DESI-like observations are compared with results from cross-correlations in \autoref{fig:gc}. We find that, in our framework, galaxy clustering provides constraints qualitatively comparable with cross-correlation. The main differences are noticeable in the shape of the 2D contours: we find stronger degeneracies in the galaxy clustering case between the cosmological parameters, while with cross-correlation some of these degeneracies are partially removed. As expected, the marked $\Omega_c h^2 - H_0$ degeneracy is present also for galaxy clustering alone, due to the fact that it derives from a measure of the matter power spectrum shape (see \cite{Berti:2022ilk}). 

\begin{figure}
	\includegraphics[width=\columnwidth]{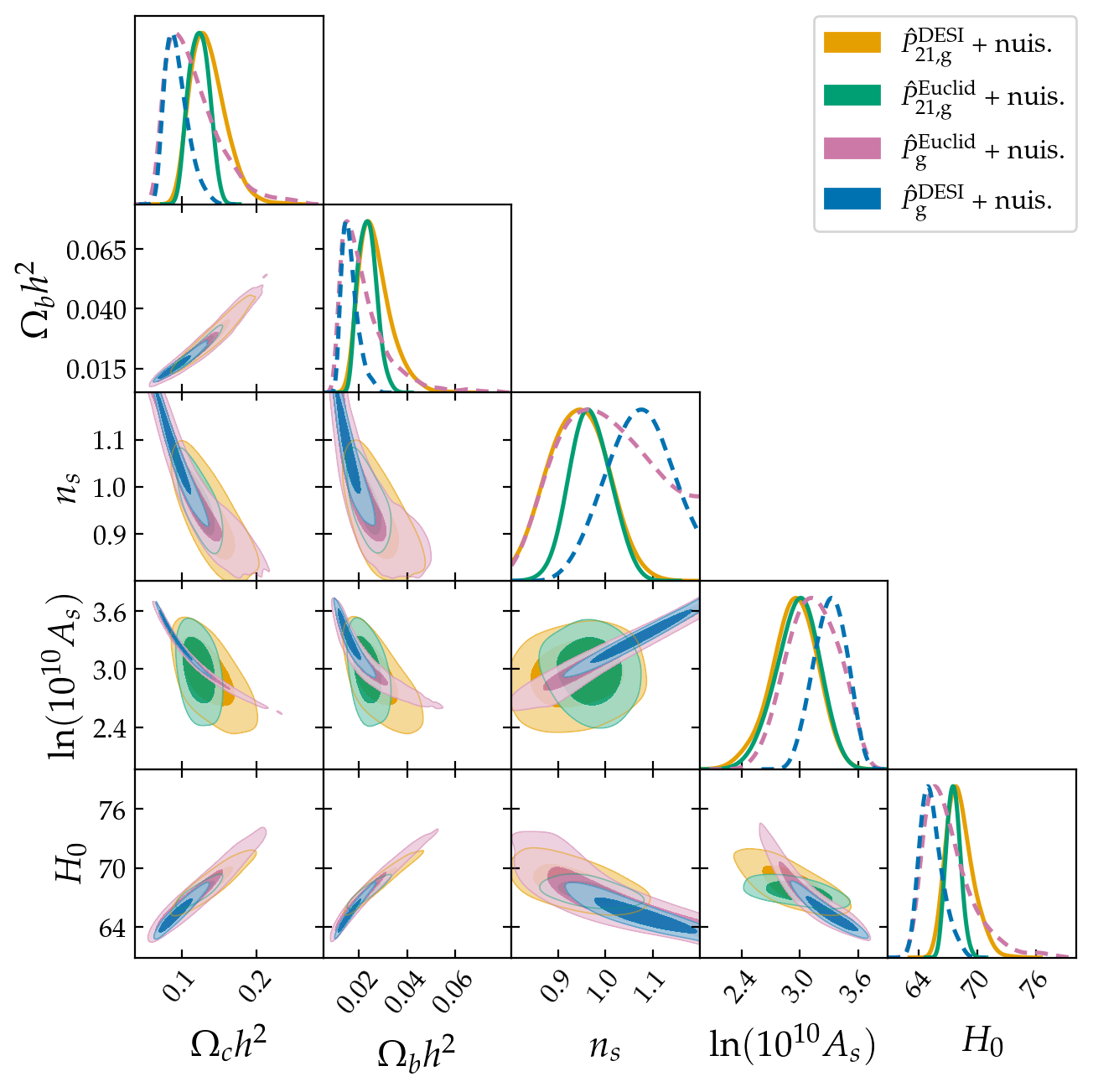}
    \caption{{Joint constraints (68\% and 95\% confidence regions) and marginalised posterior distributions on a subset of the cosmological parameters. We compare the results presented in \secref{sec:res_cross} from measurements of the 21cm and galaxy clustering cross-correlation power spectrum $\hat{P}_{21,\rm g}$ (continuous lines) with the ones from galaxy clustering auto-power spectrum $\hat{P}_{\rm g}$ (dashed lines). The label "nuis." indicates that we vary the nuisance parameters along with the cosmological ones.}}
    \label{fig:gc}
\end{figure}
}

%%%%%%%%%%%%%%%%%%%%%%%%%%%%%%%%%%%%%%%%%%%%%%%%%%

% Don't change these lines
\bsp	% typesetting comment
\label{lastpage}
\end{document}